%% file: ms.tex
\documentclass[preprint, 3p]{elsarticle}
\pdfoutput=1
\journal{JFS}
\biboptions{numbers,sort&compress}
\usepackage{lineno,hyperref}
\usepackage{newtxtext,newtxmath}
\usepackage{makeidx}         
\usepackage{graphicx}        
\usepackage{enumitem}
\usepackage{color}
\usepackage{natbib}
\usepackage{float}
\usepackage{transparent}
\usepackage{amsmath}
\usepackage{amsfonts}
\usepackage{import}
\usepackage{epsfig}
\usepackage{rotating}
\usepackage[labelformat=simple]{subcaption}

\usepackage{xspace}
\usepackage{capt-of}
\usepackage{bm}
\usepackage{epstopdf}
\usepackage{siunitx}
\usepackage{booktabs}
\usepackage{tikz}
\usetikzlibrary{arrows, shapes.geometric}

\immediate\write18{%
makeindex -s nomencl.ist -o \jobname.nls -t \jobname.nlg \jobname.nlo%
}
\usepackage{nomencl}
\makenomenclature
\usepackage{multicol}

\usepackage{etoolbox}
\renewcommand\nomgroup[1]{%
  \item[\bfseries
  \ifstrequal{#1}{S}{Superscripts}{%
  \ifstrequal{#1}{B}{Subscripts}{}}%
]}

\newcommand{\nc}{\newcommand}
\nc{\rnc}{\renewcommand}
\nc{\bs}{\boldsymbol}
\nc{\RM}[1]{\MakeUppercase{\romannumeral #1}}
\nc{\define}{:=}
\nc{\ul}{\underline}
\nc{\mm}{\boldsymbol}
\nc{\mms}{\mm}
\nc{\real}[1]{\operatorname{Re}\left\lbrace #1 \right\rbrace}
\nc{\imag}[1]{\operatorname{Im}\lbrace #1 \rbrace}
\nc{\abs}[1]{\left| #1 \right|}
\nc{\COMMENT}[1]{\textcolor{red}{#1}}
\nc{\ie}{i.e.\,\xspace}
\nc{\eg}{e.g.\,\xspace}
\nc{\cf}{cf.\,\xspace}
\nc{\etal}{et\,al.\,}
\nc{\sref}[1]{Section \ref{sec:#1}}
\nc{\aref}[1]{\ref{append:#1}}
\nc{\eref}[1]{Eq.\ (\ref{eq:#1})}
\nc{\erefs}[1]{Eqs.\ (\ref{eq:#1})}
\nc{\erefo}[1]{(\ref{eq:#1})}
\nc{\fref}[1]{Fig.\ \ref{fig:#1}}
\nc{\tref}[1]{Tab.\ \ref{tab:#1}}
\nc{\fk}{\,,}
\nc{\fp}{\,.}
\nc{\herm}{{}^{\mathrm H}}
\nc{\tra}{{}^{\mathrm T}}
\rnc{\matrix}[2]{\left[\!\!\begin{array}{#1} #2\end{array}\!\!\right]}
\nc{\ee}{\mathrm{e}}
\nc{\ii}{\mathrm{i}}
\nc{\dd}{\mathrm{d}}
\nc{\e}[2]{\begin{equation} #1 \label {eq:#2} \end{equation}}
\nc{\ea}[2]{
\begin{eqnarray}
#1 \label {eq:#2} \end{eqnarray}}
\nc{\fig}[4][tbh]{\begin{figure}[#1]
\centering
\includegraphics[width=#4\textwidth]{figs/#2}\caption{#3\label{fig:#2}}
\end{figure}}

\nc{\usensor}{\hat{u}_{\mathrm{Sensor}}}
\nc{\umodal}{\hat{u}}
\nc{\usolid}{\hat{\mm u}^\mathrm{s}}
\nc{\usolidtd}{\mm u^\mathrm{s}}
\nc{\ufluid}{\hat{\mm u}^\mathrm{a}}
\nc{\ufluidtd}{\mm u^\mathrm{a}}
\nc{\forcec}{\hat{\mm f}^\mathrm{c}}
\nc{\forcea}{\hat{\mm f}^\mathrm{a}}
\nc{\forcectd}{\mm f^\mathrm{c}}
\nc{\forceatd}{\mm f^\mathrm{a}}
\nc{\Dsnma}{D^{\mathrm{s}}(\omega^{\mathrm{nl}},\mm \psi^\mathrm{nl})}
\nc{\Danma}{D^{\mathrm{a}}(\omega^{\mathrm{nl}},\mm \psi^\mathrm{nl})}
\nc{\Dslin}{D^{\mathrm{s}}(\omega^{\mathrm{stick}},\mm \psi^{\mathrm{stick}})}
\nc{\Dalin}{D^{\mathrm{a}}(\omega^{\mathrm{stick}},\mm \psi^{\mathrm{stick}})}
\nc{\Danmaconstpsi}{D^{\mathrm{a}}(\omega^{\mathrm{nl}},\mm \psi^{\mathrm{stick}})}
\nc{\Danmaconstom}{D^{\mathrm{a}}(\omega^{\mathrm{stick}},\mm \psi^{\mathrm{nl}})}

\renewcommand{\d}[1]{\ensuremath{\operatorname{d}\!{#1}}}

\makeindex

\begin{document}

\begin{frontmatter}
\title{Development of a Fully-Coupled Harmonic Balance Method and a Refined Energy Method for the Computation of Flutter-Induced Limit Cycle Oscillations of Bladed Disks with Nonlinear Friction Contacts}
\author[addressB]{Christian Berthold}
\author[addressA]{Johann Gross}
\author[addressB]{Christian Frey}
\author[addressA]{Malte Krack}
\address[addressA]{Institute of Aircraft Propulsion Systems, University of Stuttgart, Pfaffenwaldring 6, 70569 Stuttgart, Germany\\ johann.gross@ila.uni-stuttgart.de, malte.krack@ila.uni-stuttgart.de}
\address[addressB]{Institute of Propulsion Technology, German Aerospace Center, Linder H\"ohe, 51147 Cologne, Germany \\ christian.berthold@dlr.de, christian.frey@dlr.de}
\begin{abstract}
\input{abstract}
\end{abstract}
\end{frontmatter}

\nomenclature{$\usensor$}{Displacement amplitude of sensor node in circumferential direction}
\nomenclature{$u$}{Displacement}
\nomenclature{$\mm u$}{Vector of structural coordinates}
\nomenclature{$\mm f$}{Vector of forces}
\nomenclature{$D$}{Modal damping ratio}
\nomenclature{$\mm \psi$}{Modal deflection shape}
\nomenclature{$L_\mathrm{Span}$}{Blade span}
\nomenclature{$\omega$}{Angular oscillation frequency}
\nomenclature{ND}{Nodal diameter}
\nomenclature{LCO}{Limit Cycle Oscillation}
\nomenclature{HB}{Harmonic Balance}
\nomenclature{$\mm M$}{Mass matrix}
\nomenclature{$\mm K$}{Stiffness matrix}
\nomenclature{$\Re$}{Real part}
\nomenclature{$\Im$}{Imaginary part}
\nomenclature{$\mm R$}{Residual vector}
\nomenclature{$\mm \chi$}{Finite volume mesh location}
\nomenclature{$\mm x$}{spatial coordinate}
\nomenclature{$\mm e_j$}{Unit vector pointing in the $j$-th coordinate direction}
\nomenclature{$\Delta W$}{Work per cycle}
\nomenclature{$\ii$}{imaginary unit, $\sqrt{-1}$}
\nomenclature{$t$}{time}
\nomenclature{$T$}{Time period}
\nomenclature{$G$}{Aerodynamic influence coefficient}
\nomenclature{$\mm G$}{Aerodynamic influence coefficient matrix}
\nomenclature{MAC}{Modal assurance criterion}
\nomenclature{FSI}{Fluid-Structure Interaction}
\nomenclature{$H$}{Harmonic truncation order}
\nomenclature[S]{$\hat{}$}{Fourier coefficient}
\nomenclature[S]{$\mathrm{a}$}{Aerodynamic quantity}
\nomenclature[S]{$\mathrm{s}$}{Solid quantity}
\nomenclature[S]{$\mathrm{c}$}{Friction contact}
\nomenclature[S]{$\mathrm{nl}$}{Refers to quantities of the refined energy method}
\nomenclature[S]{$\mathrm{stick}$}{Refers to fully sticking contact conditions}
\nomenclature[S]{$\mathrm{cb}$}{Refers to the Craig-Bampton model}
\nomenclature[S]{$\mathrm{fem}$}{Refers to the Finite Element model}
\nomenclature[S]{$\mathrm T$}{Transpose}
\nomenclature[S]{$\mathrm H$}{Hermitian}

\begin{multicols}{2}
\printnomenclature
\end{multicols}

\input{introduction}

\input{benchmarkmodel}
\input{energymethods}
\input{coupledsolver}
\input{results}
\input{conclusions}

\section*{Acknowledgements}
The research project was financially supported by the German Research Foundation (DFG) (funding no. 382141955) and the FVV (Research Association for Combustion Engines eV) (FVV funding no. 6013081).
The research project was performed by the Institute of Propulsion Technology (IAT) at the German Aerospace Center under the direction of Prof. Reinhard Mönig and by the Institute of Aircraft Propulsion Systems (ILA) at University of Stuttgart under the direction of Jun.-Prof. Malte Krack.
The project was conducted by an expert group under the direction of Dr. Andreas Hartung (MTU Aero Engines AG).
The authors greatfully acknowledge the support received from the FVV (Research Association for Combustion Engines eV) and the DFG (German Research Foundation).
\pagebreak
\input{appendix}
\bibliography{literature_krack,Literatur_NM}

\end{document}

%% file: abstract.tex
{\it
Flutter stability is a dominant design constraint of modern gas and steam turbines.
To further increase the feasible design space, flutter-tolerant designs are currently explored, which may undergo Limit Cycle Oscillations (LCOs) of acceptable, yet not vanishing, level.
Bounded self-excited oscillations are a priori a nonlinear phenomenon, and can thus only be explained by nonlinear interactions such as dry stick-slip friction in mechanical joints.
The currently available simulation methods for blade flutter account for nonlinear interactions, at most, in only one domain, the structure or the fluid, and assume the behavior in the other domain as linear.
In this work, we develop a fully-coupled nonlinear frequency domain method which is capable of resolving nonlinear flow and structural effects.
We demonstrate the computational performance of this method for a state-of-the-art aeroelastic model of a shrouded turbine blade row.
Besides simulating limit cycles, we predict, for the first time, the phenomenon of nonlinear instability, \ie, a situation where the equilibrium point is locally stable, but for sufficiently strong perturbation (caused \eg by an impact), the dry frictional dissipation cannot bound the flutter vibrations.
This implies that linearized theory does not necessary lead to a conservative design of turbine blades.
We show that this phenomenon is due to the nonlinear contact interactions at the tip shrouds, which cause a change of the vibrational deflection shape and frequency, which in turn leads to a loss of aeroelastic stability.
Finally, we extend the well-known energy method to capture these effects, and conclude that it provides a good approximation and is useful for initializing the fully-coupled solver.
}

%% file: introduction.tex
\section{Introduction}
Blade vibrations cause high-cycle fatigue and therefore put at risk the safe operation of gas and steam turbines.
Besides resonances with rotor-synchronous forcing, flutter is one of the most important causes for blade vibrations.
Flutter denotes a dynamically unstable interaction between an elastic structure and a surrounding flow.
In turbomachinery, flutter is mainly caused by the aerodynamic interference among the vibrating blades within a row (cascade effect) \cite{Srinivasan1997}.
The current design paradigm is to completely avoid flutter-induced vibrations.
This is achieved by linear vibration theory: The normal modes of vibration of the linearized structural model are determined, and for each relevant mode, the aerodynamic damping is computed assuming infinitesimally small vibrations.
If the sum of the (possibly negative) aerodynamic damping and the positive structural damping (estimated from experience) of each mode is positive, the design is accepted; otherwise it is viewed as \emph{aeroelastically unstable} and discarded.
Flutter avoidance has become a dominant design constraint with regard to both compressor \cite{Leichtfuss2013, May2010, Schoenenborn2012} and turbine blades \cite{Waite2014,Waite.2016}.
It is currently explored to overcome the strict requirement of flutter avoidance by flutter tolerance, where a dynamic instability is permitted as long as the flutter-induced vibrations remain bounded at an acceptably low level.
This change in design paradigm has the potential to further improve aerodynamic performance, and thus to reduce emissions and save resources.
\\
In the presence of a dynamic instability, linear theory only predicts the unbounded exponential growth of the vibrations (divergence).
A saturation and, thus, bounded self-excited oscillations are a priori a nonlinear phenomenon.
Nonlinear force-deformation relations, in principle, can have various physical origins.
For large vibrations, geometric and material nonlinearity may become important on the structural side.
Aerodynamic forces may depend nonlinearly on the vibration amplitude, \eg, in the case of separating and transonic flows.
Finally, the contact interactions occurring at mechanical joints (blade attachment in disk, underplatform friction dampers, interlocked tip shrouds, etc.\xspace) give rise to nonlinear contact forces.
This work focusses on nonlinear contact forces.
\\
In the steady state, flutter-induced vibrations typically take the form of a periodic oscillation, denoted as Limit Cycle Oscillation (LCO) \cite{M.Lassalle.2018,Martel.2014}.
This is also the case when multiple modes of vibration are aeroelastically unstable \cite{Martel.2014}.
Besides LCOs, quasi-periodic oscillations (or Limit Torus Oscillations) may also occur.
Limit Torus Oscillations are caused by the interaction between multiple unstable modes of vibration, if their frequencies satisfy a certain combination resonance condition, and nonlinearity is sufficiently strong \cite{Gross.2019}.
The methods presented in this work focus on LCOs, the generalization to Limit Torus Oscillations is left for future research.
Over a period of an LCO, the work $\Delta W^{\mathrm f}<0$ provided by aerodynamic forces cancels with that dissipated by structural forces $\Delta W^{\mathrm s}>0$, $\Delta W^{\mathrm f} + \Delta W^{\mathrm s} = 0$.
Since material damping of turbomachinery blades is usually negligible, $\Delta W^{\mathrm s}$ is dominated by dry frictional dissipation in the joints.
An LCO is characterized by its angular frequency $\omega$, deflection shape $\mm\psi$, amplitude $\hat u$ and potential higher-harmonic content.
The accurate prediction of these properties relies on a precise modeling of aerodynamic and structural forces.
For modeling of the structural forces within frictionally coupled bladed disks, we refer to a recent overview \cite{Krack2017}.
For modeling of the aerodynamic forces, several simplifying assumptions are commonly made, as discussed in the following.
\\
Firstly, it is very common to assume the aerodynamic forces as linear in the structural displacement amplitude $\hat u$.
In the amplitude range relevant for high-cycle fatigue, this assumption has been supported by experimental \cite{Nowinski.1999,Seeley2016} and numerical \cite{Corral.2008,Ren2016,Muller.2018} evidence.
\\
Secondly, it is often assumed that $\omega$ and $\mm\psi$ correspond to a certain normal vibration mode of the structure in vacuum, obtained under linearized contact boundary conditions in joints (\ie sticking contact in preloaded interfaces).
Consequently, these properties can be efficiently computed using finite elements and linear modal analysis.
The associated aerodynamic work per cycle $\Delta W^{\mathrm f}$ can then be determined for each mode of interest using computational fluid dynamics.
This leads to the \emph{conventional energy method}, where frequency and shape of the vibration are assumed as descried above, and the amplitude is determined from the requirement that supplied and dissipated works per cycle cancel each other ($\Delta W^{\mathrm f}+\Delta W^{\mathrm s} = 0$).
This strategy was followed \eg in \cite{sinh1983,sinh1985a,sinh1985b,Martel.2014}, and it is most certainly the prevailing industry practice today.
It is justified if the aerodynamic and nonlinear contact forces have negligible effect on the deflection shape $\mm\psi$ and frequency $\omega$ of the LCO.
However, the nonlinear contact boundary conditions can easily account for more than $10\%$ frequency shift and significantly alter the deflection shape \cite{Hartung.2018}, as compared to the linearized case.
Given that the reduced frequency and the deflection shape are the driving determinants for flutter, this strategy has a strictly limited range of validity.
Indeed, the sign of the aerodynamic work can be affected by the boundary conditions at the contact interfaces.
For instance, Leyes \etal \cite{Leyes.1999} associate a blade failure in an aircraft engine to a flutter instability caused by the altered boundary conditions at the tip shrouds due to fretting wear.
Similarly, Wu \etal \cite{Wu.2007} demonstrate a test case that was stable for sticking and unstable for frictionless sliding tip shroud interfaces.
\\
To account for a potential change of the deflection shape $\mms\psi$ and frequency $\omega$ with the amplitude $\hat u$, the linear aerodynamic forces can be modeled in terms of \emph{influence coefficients} in the frequency domain \cite{petr2012c,Waite2014,Martel.2014,Fuhrer2017}.
These are formulated with respect to a selected set of modes.
To predict these coefficients, a harmonic deformation in each mode is imposed (again, with infinitesimal amplitude) and the resulting aerodynamic force, more specifically its fundamental Fourier coefficient, is determined using computational fluid dynamics.
The force induced by unit deformation in mode $j$, acting on mode $i$ is the influence coefficient $G_{ij}$.
This way, a modal aerodynamic influence coefficient matrix (MAIM) $\mms G = [G_{ij}]$ is set up.
The MAIM is a complex-valued matrix; the real parts of the diagonal entries are associated with the aerodynamic stiffness and the imaginary parts with the aerodynamic damping.
The MAIM generally depends on the frequency $\omega$ of the imposed vibration.
If the actual vibration frequency is accurately known, the frequency dependence can be neglected \cite{Martel.2007}.
For cases where the relevant frequency range is large, a piecewise linear interpolation was proposed in \cite{epureanu2013}.
Just as with the conventional energy method, an advantage of the method of influence coefficients is the decoupling of computational fluid dynamics and nonlinear structural dynamics:
Once the MAIM is set up, one has a closed-form expression for the aerodynamic forces.
This can subsequently be used to compute the LCO by considering the balance with the nonlinear structural forces.
As the computational effort for setting up the MAIM grows quickly with the number of modes, the appropriate choice of a modal basis for describing structural vibration is a crucial aspect.
\\
An alternative to using influence coefficients is the multi-physical dynamic co-simulation of both structural and fluid mechanics, \ie, a \emph{fully-coupled} Fluid-Structure Interaction computation (see e.g. \cite{Berthold2018b, Im2012}).
An advantage of this approach is that it is straight-forward to account for nonlinear effects in both domains.
However, the available methods rely on numerical time integration and lead to computational efforts prohibitive for the design turbomachinery components \cite{Corral.2014,Berthold2018b}.
To overcome the high computational effort of numerical time integration, nonlinear frequency domain methods such as the Harmonic Balance (HB) method have been developed which are now popular in both structural dynamics and computational fluid dynamics.
Compared to numerical time integration, the computational effort is typically reduced by two to four orders of magnitude \cite{Kersken2017, Krack.2018a}.
The reasons for the computational advantages of HB are:
(a) HB can efficiently deal with stiff differential equations.
(b) HB avoids the costly simulation of the usually long transient.
(c) A few harmonics often suffice to resolve the periodic oscillation.
(d) Assuming symmetric behavior within the blade row, the computational domains can be reduced to a reference sector (only one blade or one passage) with appropriate cyclic symmetry boundary conditions.
First attempts to co-simulate cascade flutter in the time or frequency domain are limited to linear \cite{Ekici2012,Su2016,Cadel.2017,Gong2019} or single-degree-of-freedom nonlinear \cite{Berthold2016,Berthold2018a} structural models.
\\
As described above, the current approach to design against flutter relies on linear vibration theory.
On the one hand, this strategy is too conservative, as this excludes the design space where flutter would lead to vibrations of tolerable level.
On the other hand, the current approach is not conservative enough, as the case of nonlinear instability is ignored; i.e., a situation where the static equilibrium is aeroelastically stable, but an instability occurs for (perhaps rather small but) finite vibration amplitudes due to nonlinear effects.
There is thus a need for simulation methods that predict the nonlinear Fluid-Structure Interaction.
The purpose of this work is to develop efficient frequency-domain methods for computing flutter-induced LCOs of bladed disks, which are capable of accounting for both structural and aerodynamic nonlinear effects.
In \sref{benchmarkmodel}, the model of the considered benchmark is presented, along with the frequency-domain formulation of the governing equations in either domain.
Then, the conventional energy method is refined to take into account structural nonlinearities and their effect on the aerodynamic damping (\sref{energymethods}).
The fully-coupled solver is presented in \sref{coupledsolver}.
The computational performance of the methods is then assessed for the case of a stable LCO and an unstable LCO (stability limit) in \sref{results}.
The article ends with the conclusions in \sref{conclusions}.

%% file: benchmarkmodel.tex
\section{Benchmark and modeling\label{sec:benchmarkmodel}}

\subsection{Structural model}
A model of a low pressure turbine bladed disk with interlocked tip shrouds is considered.
The structural finite element model of one sector is shown in \fref{model_structure}; the whole bladed disk contains \num{50} sectors.
The structural modeling, briefly described in the following, represents the current state of the art, and is described in more detail in \cite{Krack2017}.
Geometric and material properties are listed in \tref{testcasedetails}.
Mistuning is known for its potential to significantly distort the modal deflection shapes and for its tendency to stabilize flutter \cite{Nowinski.1999,Martel.2007}.
On the other hand, for the case of strong inter-sector coupling, here via tip shrouds, the effect of small mistuning is known to be negligible \cite{wei1988a}.
In this work, the system is assumed as rotationally periodic; \ie, each sector has identical properties.
It is further assumed that the periodic response of the system inherits the symmetry suggested by the system; \ie, the response is assumed to take the form of a traveling wave, where each sector undergoes the same oscillation, however, with a constant time shift between neighboring sectors.
This permits to reduce the problem domain to only a reference sector with appropriate time-shift boundary conditions at the cyclic sector boundaries.
\fig[tbh]{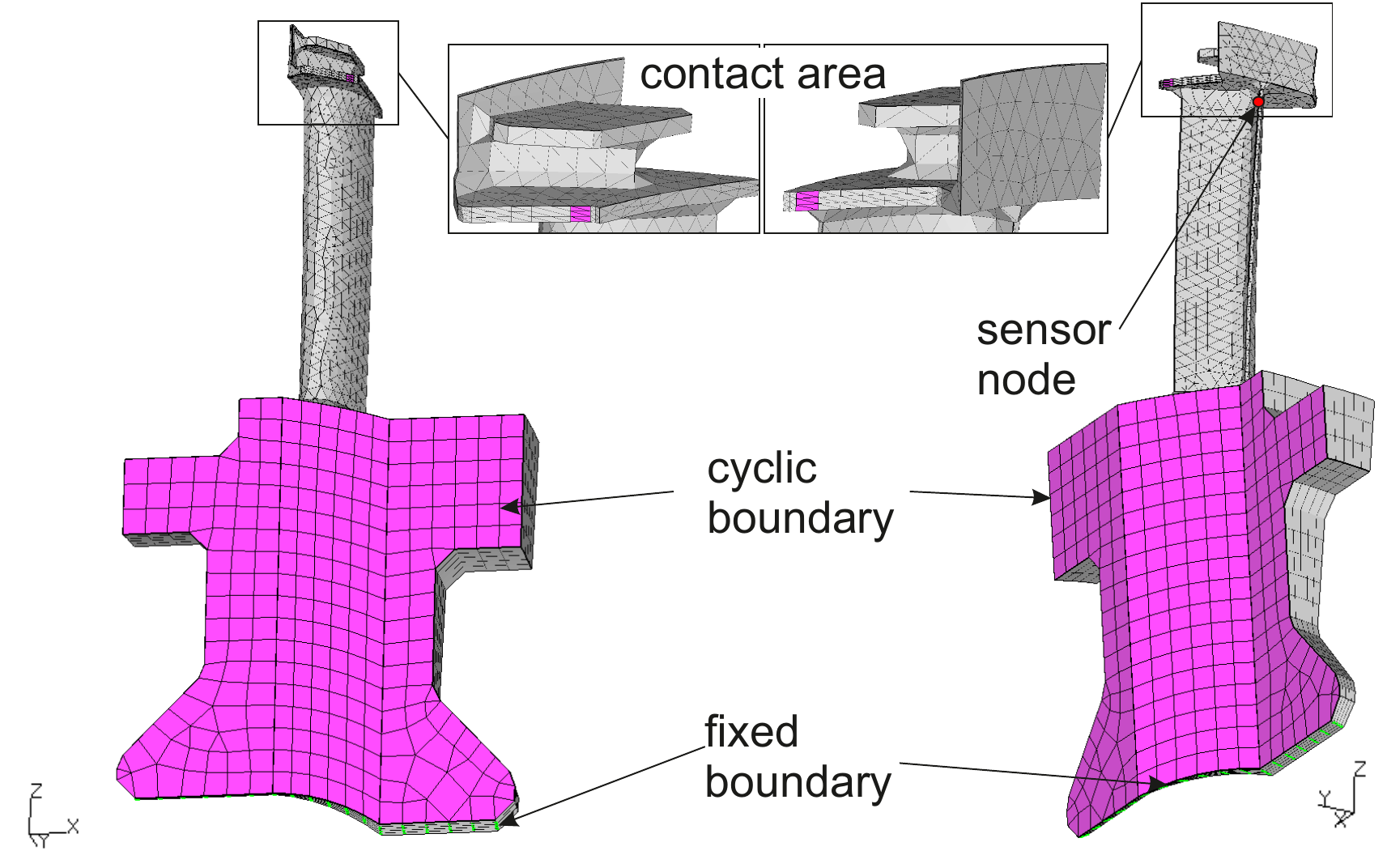}{Structural finite element model of the sector of a bladed disk (kindly provided by MTU Aero Engines AG)}{.7}
\begin{table}[h]
\centering
\begin{tabular}{rr}
\toprule
Number of blades & \num{50} \\
Blade span & \SI{86}{\milli\meter} \\
Blade chord & \SI{22}{\milli\meter} \\
Youngs modulus blade + disk &\SI{110}{\giga\pascal}\\
Poisson ratio blade + disk &\SI{0.343}{}\\
Density blade + disk &\SI{4.43e+03}{\kg\per\cubic\meter}\\
Youngs modulus shroud &\SI{210}{\giga\pascal}\\
Poisson ratio shroud &\SI{0.3}{}\\
Density shroud &\SI{7.83e+03}{\kg\per\cubic\meter}\\
Radius hub & \SI{134}{\milli\meter} \\
Radius tip & \SI{221}{\milli\meter} \\
\bottomrule
\end{tabular}
\caption{Geometric and material properties of the benchmark model.}
\label{tab:testcasedetails}
\end{table}
\\
Nonlinear dry frictional interactions are taken into account between the tip shrouds of neighboring blades.
In accordance with the cyclic symmetry assumption, the contact is defined under consideration of the time shift between the interfaces on either side of the sector.
In the tangential plane, dry friction is modeled in terms of the common elastic Coulomb law, which can be written in differential form as
\e{
\d{\mm p}_{\mathrm t} =
\left\{
\begin{array}{ll}
k_{\mathrm t}\d{\mm g}_{\mathrm t} & :~~||\mm p_{\mathrm t} + k_{\mathrm t}\d{\mm g}_{\mathrm t}|| \leq \mu p_{\mathrm n}  \\
\frac{k_{\mathrm t} \dd {\mm g}_{\mathrm t}}{||\mm p_{\mathrm t} + k_{\mathrm t}\d{\mm g}_{\mathrm t}||} & :~~||\mm p_{\mathrm t} + k_{\mathrm t}\d{\mm g}_{\mathrm t}|| > \mu p_{\mathrm n}
\end{array}
\right.
.
}{elasticcoulomblaw}
Herein, $\mm p_{\mathrm t}$ and $p_{\mathrm n}$ are tangential traction vector and normal pressure, $\mm g_{\mathrm t}$ is the tangential gap vector, and $k_{\mathrm t}$ is the tangential stiffness per area value.
A constant friction coefficient $\mu$ is considered for both static and dynamic friction.
The normal pressure $p_{\mathrm n}$ is assumed as time-constant and uniformly distributed within the contact interface.
Hence, the contact is always closed in the normal direction; liftoff is prohibited.
The specific values of $\mu p_{\mathrm n}$ and $k_{\mathrm t}$ are defined later.
The contact law in \eref{elasticcoulomblaw} is spatially discretized using six C2D3 surface elements with a total of 8 nodes (each side), conform with the underlying solid elements.
The elements and nodes of the opposing sides coincide in the cyclic sense.
The nodes are used as integration points.
Dry frictional damping in the mechanical joints is the only source of structural damping considered in this work; \ie, material damping is neglected.
\\
For convenience, a coordinate transform to the relative tangential displacements at each contact node pair is carried out (two orthogonal directions per node pair).
Assuming small vibrations around the equilibrium position, the structural elastic and inertia forces are linear in the structural displacement.
To reduce the mathematical model order, the Craig-Bampton method is applied \cite{crai1968}.
Accordingly, the vector of nodal displacements of the finite element model, $\mm u^\mathrm{fem}$ is approximated as a linear combination,
\e{
\mm u^\mathrm{fem} = \mm T^\mathrm{cb} \usolidtd \quad \text{with} \quad \mm T^\mathrm{cb} =
\begin{bmatrix}
\mm I & \mm 0\\
\mm \Psi & \mm \Phi
\end{bmatrix}\fk
}{CB}
of a set of component modes (the columns of the matrix $\mm T^{\mathrm{cb}}$) and associated coefficients (the elements of the vector $\usolidtd$).
The set of component modes is formed by static constraint modes and fixed-interface normal modes.
Here, the static constraint modes are associated with the relative tangential displacements of each contact node pair.
$\mm\Psi$ represents the static displacement of the remaining nodal degrees of freedom in response to a unit displacement of each relative tangential coordinate.
The fixed-interface normal modes are obtained by fixing all relative tangential coordinates (rigidly sticking contact); the deflection shapes associated with the lowest-frequency normal modes are collected as columns in the matrix $\mm\Phi$.
Based on the results of a convergence study, the $5$ lowest-frequency normal modes were retained.
Consequently, the dimension of $\usolidtd$ is $N^\mathrm{s}=2\times 8 + 5 = 21$.
Substituting the approximation in \eref{CB} into the equation of motion and requiring that the residual is orthogonal with respect to the component modes, yields a reduced order model of dimension $21$.
The mass and stiffness matrices, $\mm M^\mathrm{cb}$ and $\mm K^\mathrm{cb}$, in the reduced order model are related to the full-order finite element matrices via $\mm M^\mathrm{cb} = \mm T^{\mathrm{cb}\,\mathrm{H}} \mm M^\mathrm{fem} \mm T^\mathrm{cb}$ and $\mm K^\mathrm{cb} = \mm T^{\mathrm{cb}\,\mathrm{H}} \mm K^\mathrm{fem} \mm T^\mathrm{cb}$, and (aerodynamic and contact) forces are projected as $\mm f^{\mathrm{cb}} = \mm T^\mathrm{cb\,H} \mm f^\mathrm{fem}$.
The superscript $\mathrm{H}$ denotes the Hermitian (or complex-conjugate transpose).
It is emphasized that the matrix $\mm T^\mathrm{cb}$ depends on the structural properties only.
The structural vibration is now described by the reduced vector of generalized coordinates $\usolidtd$, which is the solution of the coupled fluid-structure problem.
\\
As mentioned before, we focus on LCOs; \ie, time-periodic responses.
The periodic solution is approximated with a truncated Fourier series,
\e{
\usolidtd(t) \approx \Re \left \{ \sum_{k=0}^{H^{\mathrm s}} \mathrm{e}^{\mathrm{i} k \omega t} \usolid_k \right \}\fk
}{truncatedfourierseries}
where $t$ denotes time, $\ii=\sqrt{-1}$ is the imaginary unit, $H^{\mathrm s}$ is the harmonic truncation order, $\omega$ is the fundamental oscillation frequency and $\usolid_k$ are the Fourier coefficients.
Applying the Harmonic Balance technique \cite{Krack.2019} to the reduced order model, the algebraic equation system,
\e{
\mm R^\mathrm{s}(\usolid, \omega, \forcea) := \mm S(\omega) \usolid + \forcec(\usolid) - \forcea  = \mm 0\fk
}{eomhb}
is obtained.
The Fourier coefficients are here stacked as $\usolid = [\usolid_0; \usolid_1; \dots; \usolid_H]$, where $;$ denotes vertical concatenation.
$\mm S(\omega)$ is the block diagonal dynamic stiffness matrix, which describes the linear-elastic and linear inertia forces and thus depends on $\mm M^\mathrm{cb}$ and $\mm K^\mathrm{cb}$.
$\forcec$ and $\forcea$ are the vectors of Fourier coefficients of the contact and aerodynamic forces, stacked analogous to $\usolid$.
$\forcec(\usolid)$ is computed by sampling $\usolidtd$ at equidistant time instants, to determine the steady hysteresis cycle of the contact forces in discrete time, and applying the discrete Fourier transform to determine the Fourier coefficients $\forcec$.
This procedure is well-known as alternating frequency-time scheme.

\subsection{Fluid model}
The compressible unsteady Reynolds-averaged-Navier-Stokes equations are formulated with respect to the conservative variables $\ufluidtd = [\rho, \, \rho u, \, \rho v, \, \rho w, \, \rho E]^T$ (fluid density, $x$-, $y$-, $z$- component of linear momentum, energy per volume).
The equations are closed with the $k$-$\omega$ turbulence model \cite{Wilcox1988}, the ideal gas law and the Sutherland law for the molecular viscosity.
The problem domain is discretized with a finite volume mesh containing $N_{\mathrm{cell}}=820,000$ cells per sector.
This yields a set of state variables $\ufluidtd(t)$, which are approximated analogously to \eref{truncatedfourierseries}, but with truncation order $H^{\mathrm a}$.
For each cell $j$, harmonic balance yields the equation \cite{Ashcroft.2014}:
\e{
\mm R^\mathrm{a}_{k,j}(\ufluid, \omega, \hat{\mm \chi}) := \mathrm{i}k\omega(\widehat{V \ufluidtd})_{k,j} + \widehat{\mm R}_{k,j} (\ufluid, \hat{\mm \chi}, \omega) = 0 \quad k=0,\ldots,H \quad j=1,\ldots, N_{\mathrm{cell}}\fp
}{navierstokeshb}
Herein, $\hat{\mm \chi}$ represents the mesh location, which is dynamically deformed under consideration of a Arbitrary Lagrangian-Eulerian formulation.
In accordance with the assumption of a traveling-wave-type response, only a single passage of the blade row is considered in the fluid model.
At the cyclic sector boundaries, phase lag boundary conditions are imposed.
At the inlet and outlet, two-dimensional non-reflecting boundary conditions are considered \cite{Schluess2018}.

\subsubsection{Moving Mesh Technique}
The fluid mesh must deform compatibly to the structural deformation at the blade surface.
To this end, the structural nodal displacements $\mm u^\mathrm{fem}$ are mapped onto the fluid mesh boundary $\Gamma_{\text{Blade Surface}}$,
\e{
\mm u(\mm x)^\mathrm{interpolated} = \Phi(\mm u^\mathrm{fem}, \mm x) = \Phi(\mm T^\mathrm{cb} \usolidtd, \mm x) \quad \forall \mm x \in \Gamma_{\text{Blade Surface}} \fk
}{interpolateddisplacements}
using a bilinear spatial interpolation $\Phi$ with respect to the spatial coordinate $\mm x$.
The mesh deformation is treated as a linear superposition,
\e{
\mm \chi\left(\usolidtd\right) = \mm \chi_0 + [\mm \chi_1\, \mm \chi_2 ...\mm \chi_N] \usolidtd\fk
}{deformedmeshsuperposition}
of the mesh deformations $\mm \chi_\ell$ with $N^\mathrm{s}\geq\ell\geq 1$ corresponding to a unit displacement of the $\ell$-th generalized structural coordinate, and the coordinates $\usolidtd$ used as weights.
$\mm \chi_0$ is the location of the undeformed mesh.
$\mm \chi_\ell$ with $\ell\geq 1$ are computed analogous to a linear-elastic deformation problem \cite{Voigt2010}
\ea{
\nabla \cdot \left( \mu(\mm x) \nabla \mm \chi_\ell \left(\mm x\right) \right) = 0 && \forall \mm x \in \Omega_{\text{Fluid Domain}}\fk \label{eq:movingmeshone} \\ \mm \chi_\ell(\mm x)=\Phi(\mm T^\mathrm{cb} \mm e_\ell, \mm x) && \forall \mm x \in \Gamma_{\text{Blade Surface}}\fk \label{eq:movingmeshtwo}
}{do_not_reference_me}
where the vector $\mm e_\ell$ is the $\ell$-th unit vector (containing only zeros, except for the $\ell$-th element which is equal to one), $\nabla$ denotes the gradient, $\cdot$ denotes the inner product, and $\mu(\mm x)$ corresponds to the equivalent elastic modulus.
In each cell, $\mu(\mm x)$ is made proportional to the inverse of the cell volume which makes sure that small cells, \eg in a boundary layer, are not sheared excessively.
\erefs{movingmeshone}-\erefo{movingmeshtwo} are efficiently solved with the generalized minimal residual method.
\\
Having an explicit expression for the mesh deformation, obtained in a pre-processing step, substantially reduces the computation effort compared to computing the mesh deformation simultaneously with the solution of the governing equations of the fluid.
The described procedure preserved good mesh quality throughout the amplitude range considered in this work.
By integration of the fluid pressure and traction over the blade surface, nodal forces consistent with the underlying finite elements are obtained.
Finally, these are projected onto the component modes to obtain the generalized aerodynamic forces $\mm f^\mathrm{a}$.

\subsection{Damping and Excitation}
To understand the occurrence of limit cycles and to assess their stability, it is useful to analyze the work performed by aerodynamic forces on the blade surface and the work dissipated by dry friction, per cycle of oscillation.
Therefore the damping ratio (or loss factor) is used, which is defined as \cite{ottl2007},
\e{D = \frac{\Delta W}{2 \pi E_{\mathrm p,\text{max}}}\fk}{dampingratio}
where $E_{\mathrm p,\mathrm{max}}$ is the maximum potential energy and $\Delta W$ the work per cycle.
The aerodynamic work per cycle is obtained by integrating the power due to aerodynamic forces over one oscillation period, $\Delta W^\mathrm{a} = \int_0^{2\pi/\omega}{\dot{\mm u}^{\mathrm{s}\,\mathrm T} \mm f^\mathrm{a} \dd t}$, where $\dot{\square}$ denotes derivative with respect to time $t$.
The work dissipated by dry friction is obtained analogously by $\Delta W^\mathrm{s} = \int_0^{2\pi/\omega}{\dot{\mm u}^{\mathrm{s}\,\mathrm T} \mm f^\mathrm{c} \dd t}$.
\\
A negative value of the aerodynamic damping $D^\mathrm{a}$ indicates an aerodynamic self-excitation.

\subsection{Aerodynamic damping for sticking and frictionless sliding limit cases}
\label{sec:linbehaviourlimitcases}
Elastic sticking and frictionless sliding contact conditions are linear limit cases of the elastic dry friction nonlinearity reached for sufficiently small and (asymptotically) for very large vibration amplitudes, respectively.
For very small vibration amplitudes the norm of the frictional traction will not exceed $\mu p_{\mathrm n}$.
In this case the contact elements will behave as linear springs (which corresponds to elastic sticking).
For the case of very large amplitudes the contact elements slide most of the time during the vibration cycle.
The norm of the frictional traction is limited by the finite value of $\mu p_{\mathrm n}$, and thus become negligible compared to the stresses within the body, which further increase linearly with the vibration amplitude.
Hence, the behavior for large vibrations asymptotically approaches the behavior for frictionless sliding.
\\
The normal modes of a rotationally periodic structure have the form of sinusoidal waves, standing or traveling around the circumference.
The wave number is characterized by the number of nodal diameters (ND), which are diametral lines of material points with zero deflection.
$\text{ND}=0$ means that all blades vibrate synchronously with the same amplitude (standing wave).
$\text{ND}=1$ corresponds to one sinusoidal wave around the circumference, and so on.
By convention, a positive ND refers to a wave traveling in the same direction as the rotor, while a negative ND refers to the opposite direction.
The aerodynamic damping has a characteristic dependence on the wave number and the traveling direction of a particular mode.
\fref{lcoUnstable/FlutterCurve} depicts the aerodynamic damping ratio as function of the ND for the lowest-frequency mode family, for both sticking and frictionless sliding limit cases.
The aerodynamic damping shows a considerable sensitivity to the contact boundary conditions at the tip shrouds.
The detailed view in \fref{lcoUnstable/FlutterCurveZoom} reveals that the contact boundary conditions even affect the sign of the aerodynamic damping.
The ND $-7$ mode, for instance, is aerodynamically self-excited for sticking contact conditions but positively damped for frictionless sliding contact conditions.
Consequently, a stable LCO can be expected here.
Conversely, the ND $-4$ mode receives positive aerodynamic damping for sticking, but negative damping for free sliding boundary conditions.
Thus, one can expect that this mode loses stability beyond a certain vibration level, as the nonlinear contact interactions gradually transition from the sticking to the frictionless sliding limit case.
\begin{figure}
\begin{subfigure}{.5\textwidth}
  \centering
  \includegraphics[width=.8\linewidth]{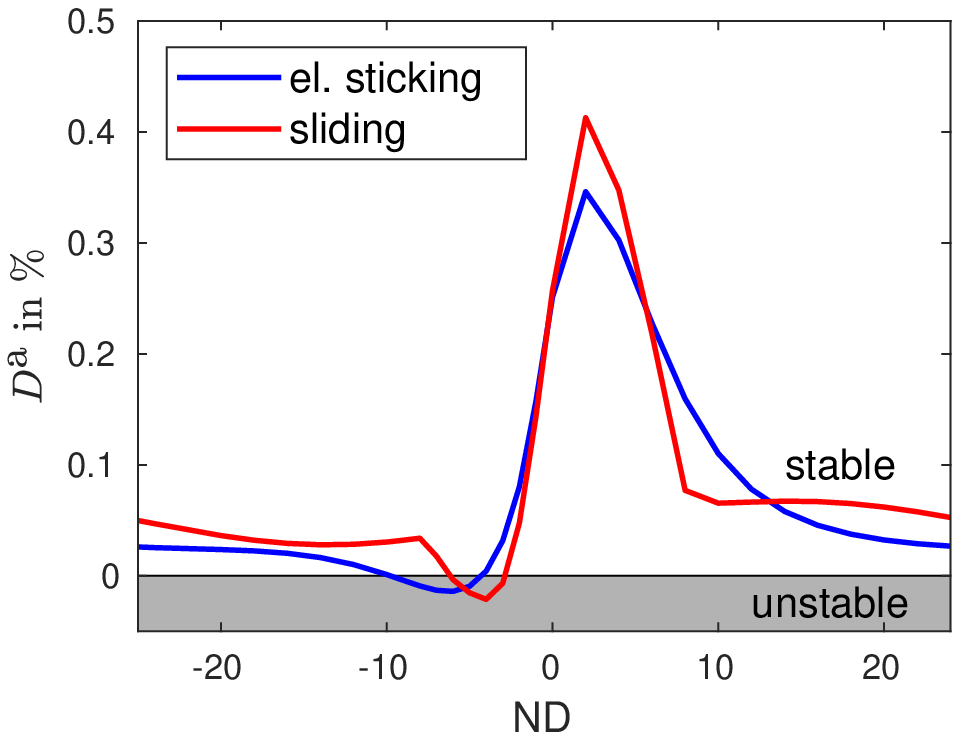}
  \caption{}
  \label{fig:lcoUnstable/FlutterCurve}
\end{subfigure}%
\begin{subfigure}{.5\textwidth}
  \centering
  \includegraphics[width=.85\linewidth]{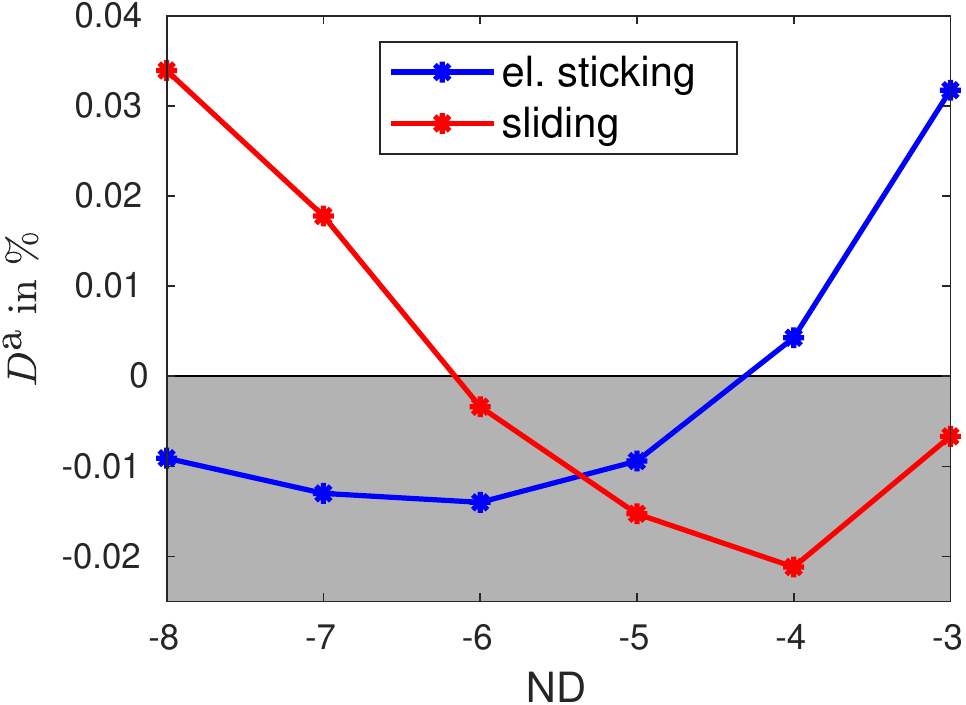}
  \caption{}
  \label{fig:lcoUnstable/FlutterCurveZoom}
\end{subfigure}
\caption{Aerodynamic damping ratio vs. ND, configuration 2: \subref{fig:lcoUnstable/FlutterCurve} Overview; \subref{fig:lcoUnstable/FlutterCurveZoom} Detail.}
\label{fig:fluttercurveUnstableLCO}
\end{figure}

\subsection{Dependence of aerodynamic forces on the structural response level}
\label{sec:linearaerodynamicforce}
\fig[h]{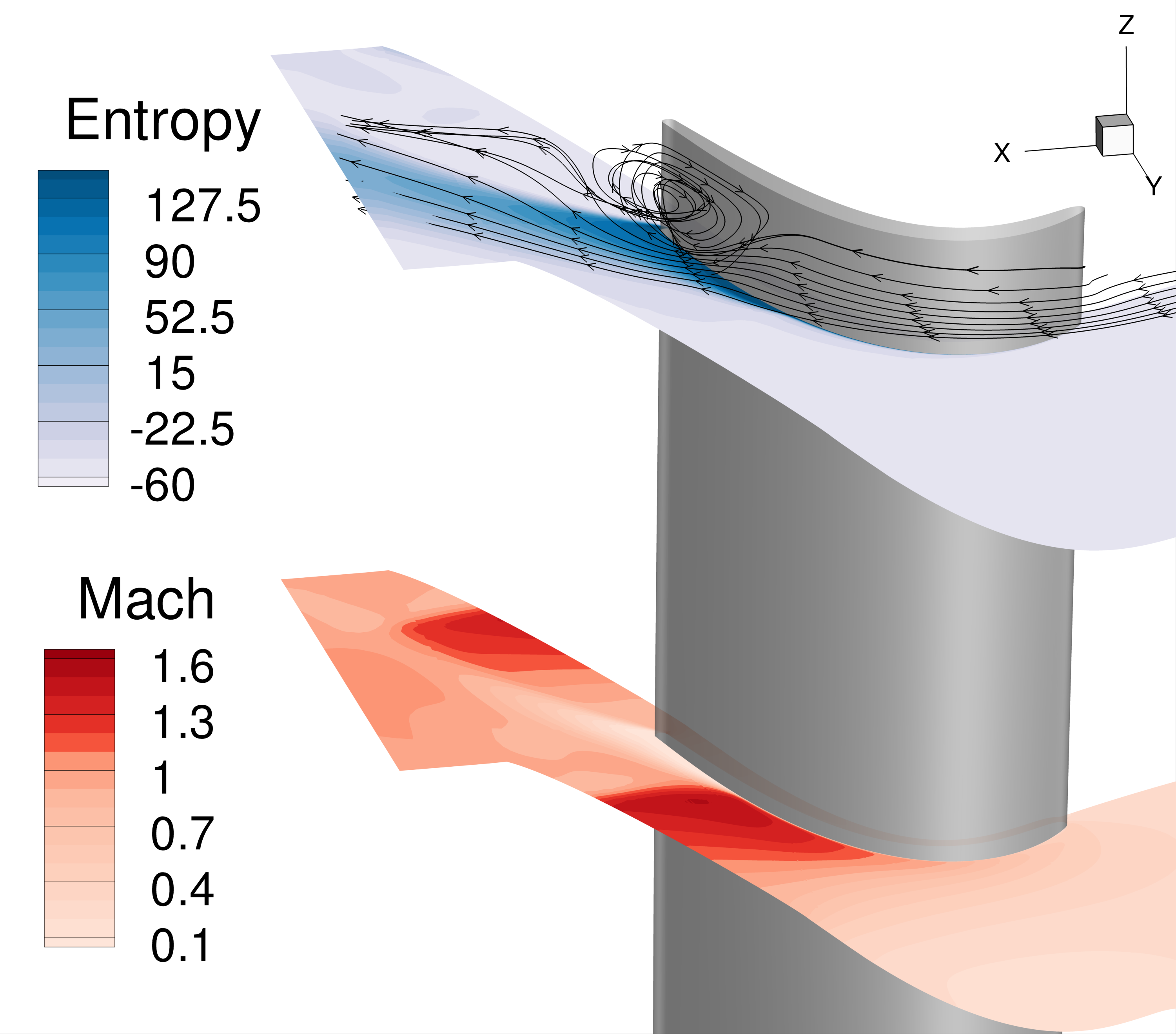}{Instantaneous flow field around the blade (mach and entropy contours).}{0.3}
\fref{FlowEntropyMach} shows an instantaneous snapshot of the cyclic symmetric flow field around the blade.
The contours are plotted at constant radius.
The lower contour shows the position of a shock and the upper contour visualizes the generation of entropy due to the flow separation in the latter part of the suction side.
The flow separation covers roughly the upper two third of the blade span.
\\
In the following, the vibrational amplitude denoted by $\usensor$ is the amplitude of the (circumferential) $Y$-component of the sensor node depicted in \fref{model_structure}.
This coordinate is chosen since the investigated mode with ND -4 deforms considerably in this direction at the selected sensor node.
For this number nodal diameters, the tip shroud displacement is relatively large, such that the displacement of the sensor node (located near the tip shroud) properly reflects the global vibration level of the blade.
In the following plots $\usensor$ is normalized by the blade span $L_\mathrm{Span}$.
\\
To analyze the amplitude-dependence of the aerodynamic forces, the aerodynamic damping is computed for different vibration amplitudes of the mode shape $\mm\psi^{\mathrm{stick}}$ and natural frequency $\omega^{\mathrm{stick}}$ obtained with sticking contact conditions.
In \fref{lcoUnstable/DampingVsLinearAmplitude} a nonlinear behavior of the aerodynamic response can be identified, however the change in damping due to the amplitude is not significant compared to the impact of the ND or the contact conditions.
For amplitudes larger than $\usensor/L_\mathrm{Span}=\num{2e-3}$ the fluid solver failed to converge. However, the computed range covers the LCO amplitudes which will later be analyzed.
\fig[tbh]{lcoUnstable/DampingVsLinearAmplitude}{Aerodynamic damping ratio vs. amplitude for ND -4 mode under sticking contact conditions, configuration 2.
}{0.4}
The modes associated with several other NDs showed similar behavior under various flow conditions for the given test case.
\\
It should be emphasized that a linear dependence of the aerodynamic forces on the vibration amplitude should not be confused with a linear dependence on the conservative flow variables.
The highly unsteady flow phenomena can certainly be captured only with nonlinear computational fluid dynamics.
Even if there is no significant direct amplitude dependence (for fixed deflection shape and frequency, $\mm\psi^{\mathrm{stick}}$, $\omega^{\mathrm{stick}}$), the nonlinear contact conditions will induce an amplitude dependent natural frequency $\omega^{\mathrm{nl}}(\hat u)$ and deflection shape $\mm\psi^{\mathrm{nl}}(\hat u)$, which in turn generates an \emph{indirect amplitude dependence} of the aerodynamic damping and stiffness.

%% file: energymethods.tex
\section{Energy methods\label{sec:energymethods}}
The basic idea behind energy methods is to first determine the frequency $\omega$ and the deflection shape $\mm\psi$, and in a second step determine the amplitude $\hat u$ by the requirement that dissipated and supplied energies per cycle cancel each other.
The result of such an analysis is illustrated in \fref{ExcitationDampingVsAmplitudePseudoDampingRatio}.
The right plot depicts $\Delta W^\mathrm{s}$ and $-\Delta W^\mathrm{a}$ as function of the amplitude $\umodal$; at the intersection points, thus, $\Delta W^\mathrm{a}+\Delta W^\mathrm{s}=0$ holds.
The corresponding amplitudes correspond to either stable or unstable limit cycles.
\begin{figure}
\begin{subfigure}{.5\textwidth}
  \centering
  \includegraphics[width=.8\linewidth]{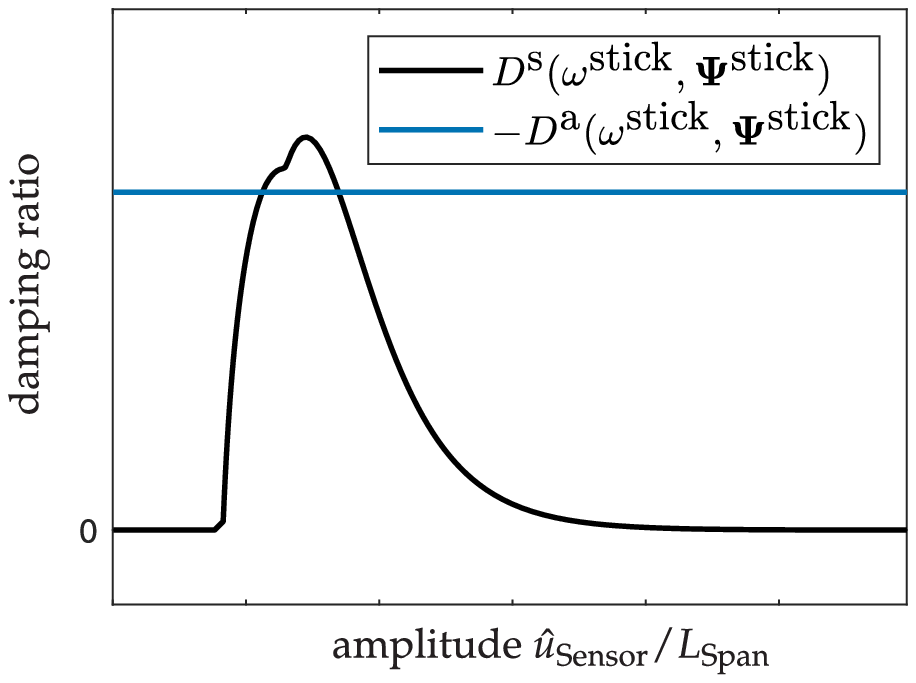}
  \caption{}
  \label{fig:ExcitationDampingVsAmplitudePseudoDampingRatio}
\end{subfigure}%
\begin{subfigure}{.5\textwidth}
  \centering
  \includegraphics[width=.8\linewidth]{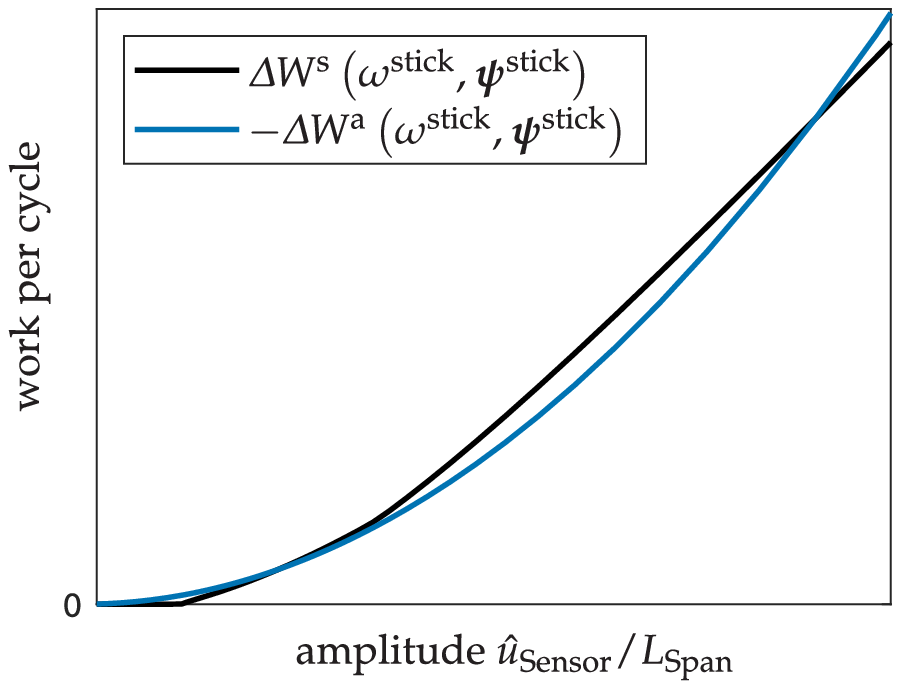}
  \caption{}
  \label{fig:ExcitationDampingVsAmplitudePseudoWork}
\end{subfigure}
\caption{Example of a stable LCO and a stability limit: \subref{fig:ExcitationDampingVsAmplitudePseudoDampingRatio} Damping ratio vs. amplitude; \subref{fig:ExcitationDampingVsAmplitudePseudoWork} Work vs. amplitude.}
\label{fig:ExcitationDampingVsAmplitudePseudo}
\end{figure}
By analyzing the dependence of $\Delta W^\mathrm{a}$ and $\Delta W^\mathrm{s}$ on the amplitude $\hat u$, one can also infer the stability of a given limit cycle:
A positive slope at the zero crossing of $\Delta W^\mathrm{a}+\Delta W^\mathrm{s}$ means that for increased amplitude, the dissipated work exceeds the supplied one.
Thus, the system returns to the limit cycle in the presence of a small perturbation; the limit cycle is stable.
If, in contrast, the slope is negative, an arbitrarily small perturbation would drive the amplitude away from this point; the limit cycle is unstable.
\\
In the example shown in \fref{ExcitationDampingVsAmplitudePseudo}, two limit cycles can be identified.
This is a typical situation under aerodynamic self-excitation and dry frictional dissipation.
Assuming that the aerodynamic forces depend linearly on the structural displacements, and the mode shape and frequency are fixed, the aerodynamic damping ratio is constant and $\Delta W^{\mathrm a}$ grows quadratically with the amplitude.
For small vibration amplitudes, the contact is sticking, and consequently $\Delta W^{\mathrm s}$ is zero.
For sufficiently large amplitudes, sliding friction occurs.
For constant normal load, the area enclosed in the force-displacement hysteresis cycle, which corresponds to $\Delta W^{\mathrm s}$, grows linearly beyond this point.
The left intersection corresponds to a stable, the right to an unstable limit cycle.
As there is no further limit cycle at higher amplitudes (for the given model), the unstable limit cycle can also be viewed as a \emph{stability limit}.
\\
The available energy methods differ by the simplifying assumptions made for the determination of $\omega$ and $\mm\psi$.

\subsection{Conventional energy method}
Traditionally, the normal modes are computed for the structure in vacuum (neglecting the aerodynamic forces) and under linearized contact conditions (here: sticking shrouds).
More specifically, $\mm \psi^\mathrm{stick}$ and $\omega^{\mathrm{stick}}$ are obtained from an eigenvalue analysis based on the finite element structural model,
$(\mm K^\mathrm{fem}+\mm K^{\mathrm{contact}} - (\omega^{\mathrm{stick}})^2 \mm M^\mathrm{fem})\mm \psi^\mathrm{stick} = \mm 0$.
For a given mode of interest, subsequently, the work dissipated by structural forces, $\Delta W^{\mathrm s}$ and that provided by aerodynamic forces, $\Delta W^{\mathrm a}$, are determined, as functions of the modal amplitude $\hat u$.
In this step, a harmonic displacement $\mm u^\mathrm{fem}(t)= \Re \left \{ \mm \psi^\mathrm{stick} \umodal e^{\ii \omega^\mathrm{stick} t} \right \}$ with amplitude $\umodal$ is imposed.
Without loss of generality, a positive real-valued amplitude $\umodal$ is assumed in the following.
\\
For blade vibration problems, it is common to assume that structural dissipation is dominated by the dry frictional contact forces in the mechanical joints, $\mm f^{\mathrm{c,fem}}$.
Indeed, material damping is neglected in this study.
Under imposed harmonic displacement $\mm u^\mathrm{fem}(t) = \mm u^\mathrm{fem}(t+T)$ with period $T=2\pi/\omega^{\mathrm{stick}}$, the steady-state contact forces will also be periodic.
The work dissipated by these forces per cycle of vibration is then
\ea{
\Delta W^{\mathrm s} (\umodal)
&=& \int\limits_{(T)} \left(\dot{\mm u}^{\mathrm{fem}}\right)^{\mathrm T} \mm f^{\mathrm{c,fem}} \dd t \nonumber\\
&=& \int\limits_{(T)} \Re\left\{\ii\omega^{\mathrm{stick}}\mm\psi^{\mathrm{stick}}\umodal\ee^{\ii\omega^{\mathrm{stick}} t}\right\}^{\mathrm T} \Re\left\{\hat{\mm f}^{\mathrm{c,fem}}_1~\ee^{\ii \omega^{\mathrm{stick}} t}\right\}\dd t\nonumber\\
&=& \int\limits_{(2\pi)} \Re\left\{\ii\mm\psi^{\mathrm{stick}}\umodal\ee^{\ii\tau}\right\}^{\mathrm T} \Re\left\{\hat{\mm f}^{\mathrm{c,fem}}_1~\ee^{\ii \tau}\right\}\dd \tau \nonumber\\
&=& \int\limits_{(2\pi)} \left(\ii \mm\psi^{\mathrm{stick}}\umodal\frac{\ee^{\ii\tau}}{2} + \overline{\ii \mm\psi^{\mathrm{stick}}\umodal}\frac{\ee^{-\ii\tau}}{2}\right)^{\mathrm T} \left( \hat{\mm f}^{\mathrm{c,fem}}_1~\frac{\ee^{\ii \tau}}2 + \overline{\hat{\mm f}}^{\mathrm{c,fem}}_1~\frac{\ee^{-\ii \tau}}2\right)\dd\tau\nonumber\\
&=& \left( -\ii\left(\mm\psi^{\mathrm{stick}}\right)^{\mathrm H} \hat{\mm f}^{\mathrm{c,fem}}_1/4 + \ii\left(\mm\psi^{\mathrm{stick}}\right)^{\mathrm T} \overline{\hat{\mm f}^{\mathrm{c,fem}}_1}/4 \right)\umodal \int\limits_{(2\pi)}\dd\tau\nonumber\\
&=& \Re\left\{-\ii\pi\left(\mm\psi^{\mathrm{stick}}\right)^{\mathrm H} \hat{\mm f}^{\mathrm{c,fem}}_1\right\}\umodal\fp
}{do_not_reference_me}
Herein, $\tau = \omega^{\mathrm{stick}}t$, $\overline{\square}$ denotes the complex conjugate and $\hat{\mm f}^{\mathrm{c,fem}}_1$ is the complex fundamental Fourier coefficient the periodic contact forces.
It should be remarked that all other harmonics of the periodic contact forces are orthogonal to the purely harmonic velocity.
Since the relative displacement is imposed, the dissipated work can be easily determined separately for each contact element, and then the sum is taken over all elements.
For given modal deflection shape $\Delta W^{\mathrm s}$ is only a function of the modal amplitude $\umodal$.
\\
For the imposed harmonic structural displacement, the flow field is simulated using computational fluid dynamics, see \eg \cite{Kersken2012,Kahl2002}.
The subsequent spatial integration over the blade surface yields the nodal aerodynamic force vector ${\mm f}^{\mathrm{a,fem}}$.
The aerodynamic work per cycle is determined in analogy to the work dissipated by dry friction,
\ea{
\Delta W^\mathrm{a}(\umodal) &=&
\Re \left\{- \ii\pi\left(\mm\psi^{\mathrm{stick}}\right)^{\mathrm H} \hat{\mm f}^{\mathrm{a,fem}}_1\right\}\umodal \nonumber\\
&=& \Re \left\{- \ii\pi\hat f^{\mathrm a}\right\}\umodal \nonumber\\
&=& \pi \Im \left\{G\right\} \umodal^2\fp
\label{eq:linearaerodynamicwork}
}{do_not_reference_me}
In the last step of this equation, it is assumed that the vibration-induced aerodynamic force is linear with respect to the modal amplitude, so that the complex amplitude of the modal aerodynamic force can be expressed as $\hat{f}^\mathrm{a} = G\umodal$, with the complex modal aerodynamic influence coefficient $G$.
Like the work dissipated by dry friction, the aerodynamic work per cycle is a function of $\umodal$ only for given modal deflection shape and natural frequency.

\subsection{Refined energy method}
\label{sec:refinedenergymethod}
In the following, we propose a refinement of the conventional energy method, which takes into account the nonlinear amplitude-dependence of mode shape, $\mm \psi^\mathrm{nl}(\umodal)$, and natural frequency, $\omega^{\mathrm{nl}}(\umodal)$.
To determine these, a nonlinear modal analysis is carried out in accordance with the Extended Periodic Motion Concept \cite{Krack.2015a}.
Again, Harmonic Balance is used to compute the periodic oscillations.
This yields the algebraic equation system,
\e{\mm S_k(\omega^{\mathrm{nl}}) \usolid_k + \forcec_k(\usolid) = 2D^\mathrm{s}\omega^{\mathrm{nl}} (\mathrm{i} k \omega^{\mathrm{nl}})\mm M^\mathrm{cb}_k \usolid_k~~k=0,1,\ldots,H. }{nmahb}
The term on the right hand side represents the mass-proportional excitation term in accordance with the definition in \cite{Krack.2015a}.
Here, the effect of the flow is still neglected (as in the conventional energy method), but the nonlinear boundary conditions at the contact interfaces are considered.
The nonlinear modal analysis starts with a small amplitude in the linear regime in which only sticking (but no sliding) friction occurs.
By numerical path continuation, the amplitude-dependent mode shapes $\mm \psi^\mathrm{nl}(\umodal)$ and natural frequency $\omega^\mathrm{nl}(\umodal)$ are obtained.
The mode shape is mass normalized, \ie, $\usolid_1 = \mm \psi^\mathrm{nl}\umodal$ with $\left(\mm \psi^\mathrm{nl}\right)\herm \mm M^{\mathrm{cb}}\mm\psi^{\mathrm{nl}}$ as in the linear case.
The modal deflection shape in the nodal displacement coordinates of the initial finite element is recovered according to $\hat{\mm u}^\mathrm{fem} = \mm T^\mathrm{cb} \mm \psi^\mathrm{nl}(\umodal) \, \umodal$.
The work dissipated by dry friction immediately follows from \eref{dampingratio}.
\\
To account for the change of the mode shape, the aerodynamic force $\forcea$ is approximated using the aerodynamic influence coefficient matrix $\mm G$ with respect to the component modes,
\e{
\forcea(\umodal) = \mm G\left(\omega^\mathrm{nl}\left(\umodal\right)\right) \mm \psi^\mathrm{nl}(\umodal) \, \umodal\fp
}{aeroforcenma}
$\mm G$ is set up column-wise, $\mm G = [\mm G_1\, \ldots\, \mm G_N]$, where $\mm G_i = \mm T^{\mathrm{cb\,H}} \hat{\mm f}^\mathrm{a,fem}_1\left(\mm T^\mathrm{cb} \mm e_i\right)$ for $i=1,\ldots,N$.
Thus, the fundamental Fourier coefficient of the aerodynamic forces must be determined, obtained under a harmonic structural displacement in the form of every component mode (index $i$).
To account for the change of the natural frequency, $\mm G$ is linearly interpolated as $\mm G(\omega) = \mm G^\mathrm{slip} + (\mm G^\mathrm{stick} - \mm G^\mathrm{slip}) (\omega - \omega^\mathrm{slip})/(\omega^\mathrm{stick} - \omega^\mathrm{slip})$.
Herein, $\omega^\mathrm{slip}$ and $\omega^\mathrm{stick}$ are the natural frequencies under the frictionless sliding and sticking linear limit cases,
and $\mm G^\mathrm{slip}$, $\mm G^\mathrm{stick}$ are obtained by imposing the structural displacement with the corresponding frequency.
The aerodynamic work per cycle is then given by
\e{
\Delta W^\mathrm{a}(\umodal) = 
\Re \left\{ -\mathrm{i} \pi \left(\mm \psi^{\mathrm{nl}}(\umodal)\right)\herm \mm G \left( \omega^\mathrm{nl}(\umodal) \right) \mm \psi^\mathrm{nl}(\umodal) \right\}\umodal^2\fp
}{waeronma}
An alternative to using the influence coefficient matrix in \eref{aeroforcenma}, the aerodynamic forces can be computed by imposing the deflection shape $\mm\psi^\mathrm{nl}(\umodal)$ and frequency $\omega^\mathrm{nl}(\umodal)$ directly, for specific values of the amplitude $\umodal$.
In preliminary investigations, it was found that the results of both alternatives are in very good agreement in the relevant amplitude and frequency range, in spite of the relatively large frequency shift by about $19\%$.
It should be remarked that the situation might be different for another benchmark.
The analysis of $\mm G(\omega)$ can provide deeper insight into the aerodynamic interaction of certain modes of vibration.
The direct evaluation of $\hat{\mm f}^{\mathrm a}_1$ using $\mm\psi^\mathrm{nl}(\umodal)$ and $\omega^\mathrm{nl}(\umodal)$ can indeed be more computationally feasible, as discussed later, and is not a priori limited to linear amplitude- and frequency-dependence of the aerodynamic forces. 

%% file: coupledsolver.tex
\section{Fully-coupled solver}
\label{sec:coupledsolver}
The algorithm of the fully-coupled solver is illustrated in \fref{coupledsolverscheme}.
A serial coupling strategy is followed, where the structural and the fluid subproblems are solved in an alternating way.
An initial guess for the frequency $\omega_{(0)}$ and the structural deformation $\usolid_{(0)}$ (in terms of Fourier coefficients of the generalized coordinates) is required, as defined later.
The outer iteration loop starts by updating the flow mesh deformation in accordance with the periodic structural displacement, as described in \sref{benchmarkmodel}.
Next, the harmonic balance residual of the fluid domain (\eref{navierstokeshb}) is minimized by iteratively improving the flow variables $\hat{\mm u}^{\mathrm a}$.
In this step, the frequency, $\omega$, and the Fourier coefficients of the structural deformation, $\usolid$, are held constant.
Once the residual norm falls below a specified tolerance, the aerodynamic force is determined.
Subsequently, the harmonic balance residual of the structural domain (\eref{eomhb}) is minimized by iteratively improving both the frequency $\omega$ and the structural variables $\usolid$.
In this step, the aerodynamic force, $\forcea$, is considered as input.
As the residuals in both domains are always reduced below the specified tolerance, the current estimate is considered as solution of the coupled problem if it does not change between two iterations (according to a given tolerance).
\\
An important benefit of the serial coupling strategy is that it is not intrusive, in the sense that existing codes for either domain can be used, and individual single-domain solvers can be employed that are optimized for the individual subproblems.
For the structural model, a Newton-like solver with analytical gradients is used, as implemented in the open source Harmonic Balance tool NLvib \cite{Krack.2019}.
The fluid model is implemented in the tool TRACE \cite{Frey2014,Ashcroft2014}, where the Harmonic Balance equations are solved with a pseudo time stepping scheme.
\\
In contrast to energy methods, it is straight-forward to account for all nonlinear effects that can be modeled with the underlying fluid and structure solvers.
Hence, the fully-coupled solver is more versatile and has the potential for a much higher prediction accuracy.
It should be emphasized that the vibrational frequency is not a priori known and considered as part of the solution.
The more complicated situation of a {\it not} locked-in vibration of the coupled system (as it can occur in case of vortex shedding or acoustic resonance), however, must still be investigated and is not part of the paper.
\\
A key to the computational efficiency of the coupled solver is that the data exchanged between the respective single-domain solvers does not require significant amount of memory.
Only Fourier coefficients of generalized coordinates and forces plus the oscillation frequency are exchanged, so only $(2H+1)N+1$ with $H=\min\left(H^{\mathrm s},H^{\mathrm a}\right)$ real numbers must be exchanged.
In this work, with $H^{\mathrm s}=1$, $H^{\mathrm a}=2$ and $N=21$, $64$ real numbers are exchanged between the single-domain solvers.
This contrasts the common multi-physics simulation tools, which exchange displacements and forces at all surface nodes and possibly many time levels.
In the following subsections, a few crucial aspects for the computational efficiency and robustness of the full-coupled solver are addressed.
\input{figs/coupledsolverscheme}

\subsection{Treatment of unknown frequency and exclusion of trivial solution}
\label{sec:avoidancetrivialsolution}
As the system is autonomous, a given periodic solution can be time-shifted arbitrarily and will still remain a solution.
In other words, the phase of the solution is arbitrary in the autonomous case.
To achieve local uniqueness, it is common practice to impose a phase constraint.
Taking into account that the oscillation frequency is an additional unknown in the autonomous case, this phase constraint is also necessary to make the number of equations equal to the number of unknowns.
Typically, the real or imaginary part of the fundamental Fourier coefficient of a specific coordinate is commonly set to zero.
While this constrains the phase, it does not exclude the trivial solution (static equilibrium, $\hat{\mm u}^{\mathrm s}=\mm 0$).
Without appropriate countermeasures, therefore, the structural solver has the tendency to run into the static equilibrium.
To overcome this, we propose a different phase constraint:
We set the real part of the fundamental Fourier coefficient of a selected generalized coordinate to a suitable \emph{nonzero} value (\fref{avoidancetrivialsolution}).
Since the solver can still adjust the imaginary part, this does not constrain the physical solution (except if the prescribed real part exceeds the magnitude, which can be easily checked upon solving).
We selected the generalized coordinate associated with the dominant fixed-interface mode.
This is expected to provide great numerical performance, as this coordinate characterizes the motion of the structure very well and is not likely to suddenly flip or significantly change its phase relative to the rest of the coordinates during the solution process.
It is important to emphasize that the described procedure has no impact on the physical solution but is actually necessary to obtain a well posed problem.
\input{figs/avoidancetrivialsolution}

\subsection{Linearization of flow force \label{sec:flowlin}}
Within the structural subproblem, the aerodynamic force $\forcea$ is modeled as linear with respect to the vibration amplitude, but always strictly consistent with the force $\forcea_{(m)}$ output by the previous run of the fluid solver.
This ensures that the converged solution is independent of the assumed linearization.
However, the linearization greatly speeds up the overall convergence of the coupled solver as the assumed linearity is at least locally a reasonable approximation.
As mentioned in the introduction, linearity with respect to the vibration amplitude is well-supported by experimental and numerical evidence, including the results presented in this study (\fref{lcoUnstable/DampingVsLinearAmplitude}).
Two variants of the described linearization are presented in the following.

\subsubsection{Frequency-dependent influence coefficient matrix}
If the fully-populated influence coefficient matrix is available the aerodynamic force $\forcea$ in \eref{eomhb} can be split as
\e{
\hat{\mm f}^{\mathrm a}_1(\usolid, \omega) = \hat{\mm f}^{\mathrm a}_{1,\,(m)} + \mm G\left(\omega\right) \hat{\mm u}^{\mathrm s}_1 - \mm G \left(\omega_{(m-1)} \right) \hat{\mm u}^{\mathrm s}_{1,\,(m-1)} \quad \hat{\mm f}^{\mathrm a}_k = \hat{\mm f}^{\mathrm a}_{k,\,(m)}\,\, \forall k\neq 1\fp
}{linearizationgmatrix}
Note that $\forcea_{(m)}$ is obtained by imposing $\usolid_{(m-1)}$ and $\omega_{(m-1)}$ in the previous run of the fluid solver (\fref{coupledsolverscheme}).
$\mm G(\omega) \usolid$ can be viewed as the part of the aerodynamic forces explainable by the influence coefficient matrix.

\subsubsection{Frequency-independent influence coefficient matrix}
A simplified variant of the previous approach is to neglect the frequency dependance of $\mm G$.
This seems justified if the change in frequency due to nonlinear effects is small.
The simplification halves the computational effort in the pre-processing, since $\mm G$ does not have to be computed at two frequencies in order to obtain an interpolation with respect to $\omega$.

\subsubsection{Dominating-mode linearization}
The disadvantage of the above method is that the computation of a fully-populated influence coefficient matrix is relatively expensive.
Another possibility is to linearize the aerodynamic force, in an ad-hoc way, as
\e{
\forcea = \forcea_{(m)} \frac{\mm e_i^\mathrm{T} \mm u^{\mathrm s}_1}{\mm e_i^\mathrm{T} \mm u^{\mathrm s}_{1,\,(m-1)}}.
}{linearizationsingledof}
Recall that $\mm e_i$ is the unit vector whose $i$-th element is 1 and the rest are zeros.
$\mm e_i^T \usolid_1$ thus extracts the $i$-th element of $\usolid_1$, which is, again, the fundamental Fourier coefficient of the dominant fixed-interface mode.
This method will perform well if the change of the mode shape and thus the change of $\forcea$ between two iterations of the coupled solver is moderate.

\subsection{Initialization}
A good initialization of the fully-coupled solver is important.
We used the results from the refined energy method
\e{
\usolid_{(0)} = \mm \psi^\mathrm{nl}(\hat{u}^\mathrm{lco}) \, \hat{u}^\mathrm{lco}, \qquad
\omega_{(0)} = \omega^\mathrm{nl}(\hat{u}^\mathrm{lco})\fk
}{}
where $\hat{u}^\mathrm{lco}$ is the modal amplitude of the limit cycle predicted by the refined energy method.
The phase of $\usolid_{(0)}$ is modified in order to meet the phase constraint illustrated in \fref{avoidancetrivialsolution}. 

%% file: figs/coupledsolverscheme.tex
\begin{figure}
\centering
\tikzstyle{block} = [rectangle, draw, text width=14em, text centered,  minimum height=2.5em, rounded corners=0.75em]
\tikzstyle{invisibleblock} = [rectangle, text width=6em,  minimum height=2.5em]
\tikzstyle{decision} = [diamond, aspect=2, draw, text width=9.0em, text centered,  minimum height=2.5em]
\tikzstyle{post} = [->,shorten >=1pt,>=stealth',thick]
\begin{tikzpicture}[node distance = 16em, auto]
	\def\yshiftnode{8}
	
    \node [block] (structure) {\textbf{structure}\\ iterate on $\usolid$, $\omega$ until $ \left|\left| \mm R^\mathrm{s} \left( \usolid, \omega, \forcea_{(m)} \right) \right|\right| < \epsilon_s$};
    
    \node [decision, below of=structure,yshift=\yshiftnode em] (convergencetest) {$ \left| \omega_{(m)}-\omega_{(m-1)} \right| < \epsilon_\omega$\\$ \left|\left| \usolid_{(m)}-\usolid_{(m-1)}\right|\right|<\epsilon_u$};
    
    \node [block, below of=convergencetest,yshift=\yshiftnode em] (flowmesh) {\textbf{flow mesh}\\ $\hat{\mm \chi}_{(m)} = [\mm \chi_1 \, \mm \chi_2 \dots \mm \chi_N] \usolid_{(m)}$\\ $m \leftarrow m + 1$};
    
    \node [block, below of=flowmesh,yshift=\yshiftnode em] (flow) {\textbf{flow}\\ iterate on $\ufluid$ until $ \left|\left| \mm R^\mathrm{a} \left( \ufluid, \omega_{(m-1)}, \hat{\mm \chi}_{(m-1)} \right) \right|\right| < \epsilon_a$};
    
    \node [block, left of=flowmesh] (aeroforce) {\textbf{aerodynamic force}\\ $\forcea_{(m)} = \forcea \left( \ufluid_{(m-1)},\omega_{(m-1)} \right) $};
    
    \node [invisibleblock, right of=convergencetest] (solution) {\textbf{solved}};
    
    \node [invisibleblock, right of=flowmesh] (initialization) {\textbf{initialization}\\ $\usolid_{(0)}, \omega_{(0)}$};
    
    \draw [post] (structure) -- (convergencetest);
    \draw [post] (convergencetest) -- (flowmesh) node[right, midway] {no};
    \draw [post] (flowmesh) -- (flow);
    \draw [post] (convergencetest) -- (solution) node[above, midway] {yes};
    \draw [post] (initialization) -- (flowmesh);
    \draw [post,rounded corners=5pt] (flow)-|(aeroforce);
    \draw [post,rounded corners=5pt] (aeroforce)|-(structure); 
\end{tikzpicture}
\caption{The algorithm scheme for the FSI frequency domain solver.}
\label{fig:coupledsolverscheme}
\end{figure}
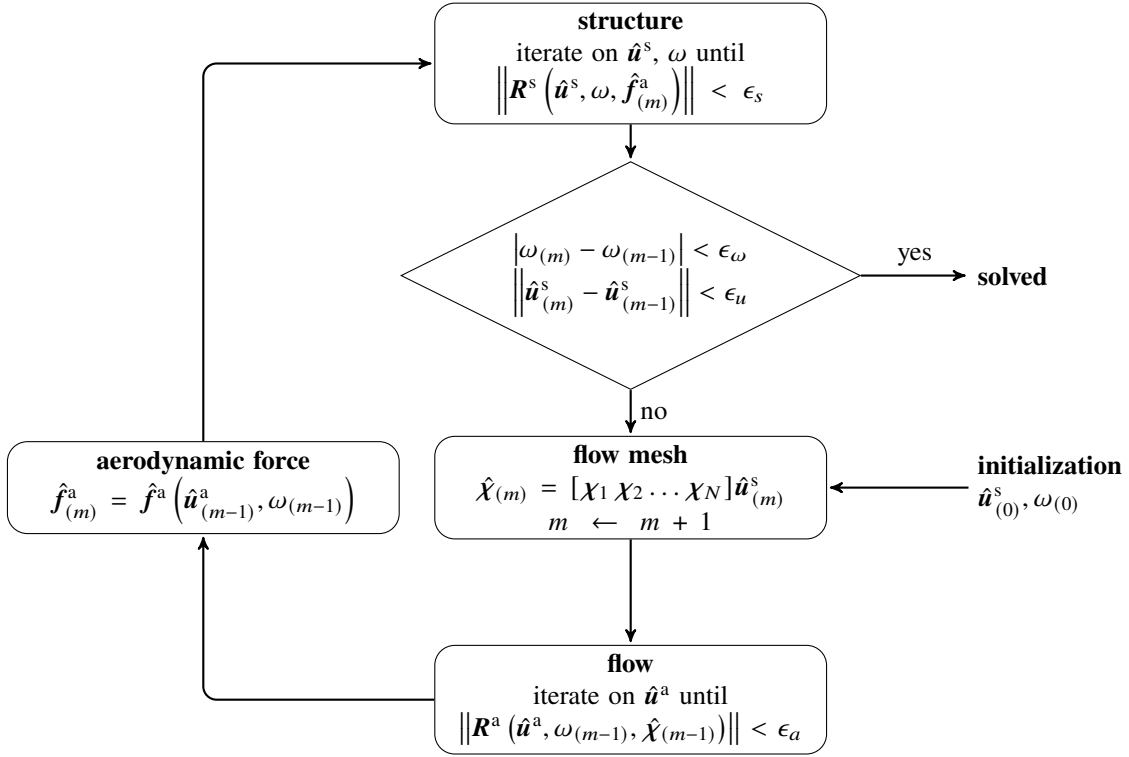

%% file: figs/avoidancetrivialsolution.tex
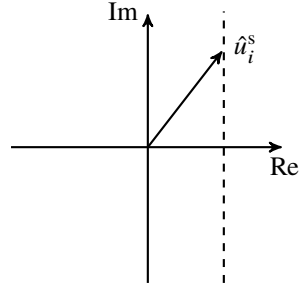
\begin{figure}
\centering
\tikzstyle{arrow} = [->,shorten >=1pt,>=stealth',thick]
\tikzstyle{constvalue} = [thick, dashed]
\begin{tikzpicture}[auto]
	\def\axislength{1.8}
	\def\constantre{1.0}
    \draw [arrow] (-\axislength,0) -- ++(2*\axislength,0) node[below] {Re};
    \draw [arrow] (0,-\axislength) -- ++(0,2*\axislength) node[left] {Im};
    \draw [arrow] (0,0) -- ++(\constantre,1.3) node[right] {$\textstyle\hat{u}^\mathrm{s}_i$};
    \draw [constvalue] (\constantre,-\axislength) -- ++(0,2*\axislength);
\end{tikzpicture}
\caption{Non-trivial phase constraint to avoid trivial solution.}
\label{fig:avoidancetrivialsolution}
\end{figure} 

%% file: results.tex
\section{Results\label{sec:results}}

\subsection{Benchmark configurations}
If multiple ND modes are aerodynamically self-excited, it is expected that each of these gives rise to a limit cycle \cite{Martel.2014}.
The larger the aerodynamic self-excitation ($-D^{\mathrm a}$) of a given mode, the larger is the corresponding basin of attraction, \ie, the set of initial conditions from which the trajectories approach the respective limit cycle.
As a representative example, we analyze the fundamental mode with ND $-4$, which also has the highest aerodynamic self-excitation in the frictionless sliding limit case shown in \fref{fluttercurveUnstableLCO}.
\begin{table}[ht]
\centering
\begin{tabular}[t]{rrr}
\toprule
configuration: & 1 & 2\\
\midrule
inlet total pressure & \SI{1.9e+05}{\pascal} & \SI{1.9e+05}{\pascal} \\
static pressure at outflow & \SI{9.0e+04}{\pascal} & \SI{7.8e+04}{\pascal} \\
contact area & $\SI{4.55}{\milli\meter\squared}$ & $\SI{4.55}{\milli\meter\squared}$ \\
contact stiffness $k_{\mathrm t}$ & \SI{2.19}{\newton\per\cubic\milli\meter}&\SI{21.95}{\newton\per\cubic\milli\meter}\\
limit frictional traction $\mu p_{\mathrm n}$ & \SI{0.22}{\newton\per\milli\meter\squared} & \SI{6.58e-4}{\newton\per\milli\meter\squared} \\
$D^\mathrm{a}$ (stick) & \num{4.3e-5} & \num{2.1e-4} \\
$D^\mathrm{a}$ (slip)  & * & \num{-1.9e-4} \\
natural frequency (stick) & \SI{437}{\hertz} & \SI{520}{\hertz} \\
natural frequency (slip)  & \SI{418}{\hertz} & \SI{418}{\hertz} \\
\bottomrule
\end{tabular}
\caption{Parameters of the two benchmark configurations and corresponding linear modal properties. *The aerodynamic damping $D^\mathrm{a}$ (slip) of configuration 1 (used for the stable limit cycle) was not computed since it is not relevant.}
\label{tab:testcaseconfigurations}
\end{table}
\\
Two configurations are studied in the following, which differ in terms of contact properties and static outflow pressure, as listed in \tref{testcaseconfigurations}.
During a long service life, the normal preload in the interlocked shrouds can decrease significantly due to creep and wear.
This is the motivation to set different normal preloads in the two configurations.
In this sense, configuration 1 corresponds to the \emph{initially built setting}, while configuration 2 corresponds to a \emph{setting after a long service life}.
First the expected and well-known stable limit cycle is analyzed for configuration 1.
As discussed in \sref{benchmarkmodel}, the ND $-4$ mode is stable under sticking contact, but unstable under frictionless sliding contact in configuration 2.
This suggests that there is an unstable limit cycle (stability limit), which will be analyzed subsequently.
\\
For the results presented in the following, a single harmonic was found sufficient for the Harmonic Balance equation of the structural subproblem ($H^{\mathrm s}=1$), while two harmonics were needed to properly resolve the dynamic processes in the fluid domain ($H^{\mathrm f}=2$).

\subsection{Stable Limit Cycle (configuration 1)}

\subsubsection{Comparison of the results obtained by energy methods and fully-coupled solver}
\fref{lcoStableNMA} shows the structural and aerodynamic damping ratios as function of the vibration amplitude.
For sufficiently small vibration amplitudes, the contact is always sticking, and thus the structure behaves linear and there is no frictional dissipation ($D^\mathrm{s}=0$).
As soon as the vibration amplitude is sufficiently large (here beyond $\left|\hat u_{\mathrm{Sensor}}\right|\approx 0.185\%\, L_{\mathrm{Span}}$), dissipative sliding friction starts to occur, which leads to an increase of $D^\mathrm{s}$.
For the given contact stiffness and area, a maximum structural damping ratio of about $4\%$ is reached.
The aerodynamic damping ratio $D^{\mathrm a}$ is negative, has a relatively small magnitude and does not considerably depend on the amplitude, as compared to $D^\mathrm{s}$.
Recall that a limit cycle requires $D^{\mathrm s} + D^{\mathrm a}=0$.
Hence, a limit cycle occurs at the intersections in \fref{lcoStable/ExcitationDampingVsAmplitude} where $D^{\mathrm s} = - D^{\mathrm a}$.
Because of the relatively small magnitude of $D^{\mathrm a}$, the aerodynamic self-excitation is cancelled by friction damping already just shortly beyond the onset of sliding friction.
Consequently, the oscillation frequency at the limit cycle is very close to that of the sticking limit case.
Indeed, the difference in frequency is only $\SI{0.003}{\percent}$ here.
Therefore, the frequency-dependence in the refined energy methods was neglected in this case.
In \tref{datastableLCO} the results of the energy methods and the fully-coupled solver are summarized in terms of amplitude and frequency of the stable limit cycle.
It can be concluded that all methods are in very good agreement for this particular case.
For completeness, also the change of the vibrational deflection shape is assessed by computing the $MAC$ (modal assurance criterion) defined as the correlation measure
\e{
MAC(\mm \psi_1,\mm\psi_2)=\frac{|\mm \psi_1^{\mathrm H} \mm \psi_2|^2}{(\mm \psi_1^{\mathrm H} \mm \psi_1) (\mm \psi_2^{\mathrm H} \mm \psi_2)}\fp
}
{mac}
The $MAC$ values shown in Table \ref{tab:datastableLCO} are all very close to unity.
Thus, it is concluded that the deflection shape at the stable limit cycle is practically identical in all methods and agrees well with the modal deflection shape obtained for sticking contact conditions.
\begin{figure}
\begin{subfigure}{.5\textwidth}
  \centering
  \includegraphics[width=.8\linewidth]{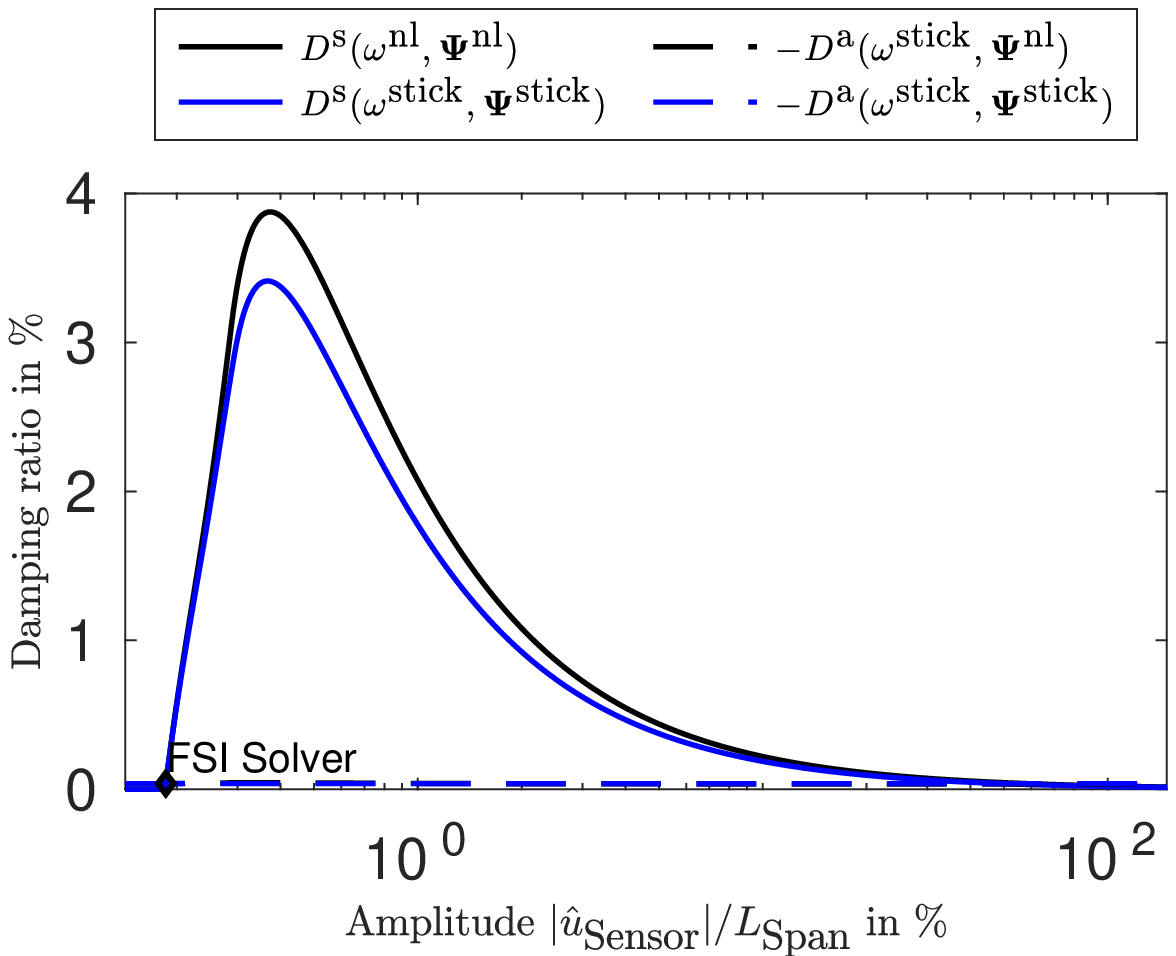}
  \caption{}
  \label{fig:lcoStable/ExcitationDampingVsAmplitude}
\end{subfigure}%
\begin{subfigure}{.5\textwidth}
  \centering
  \includegraphics[width=.8\linewidth]{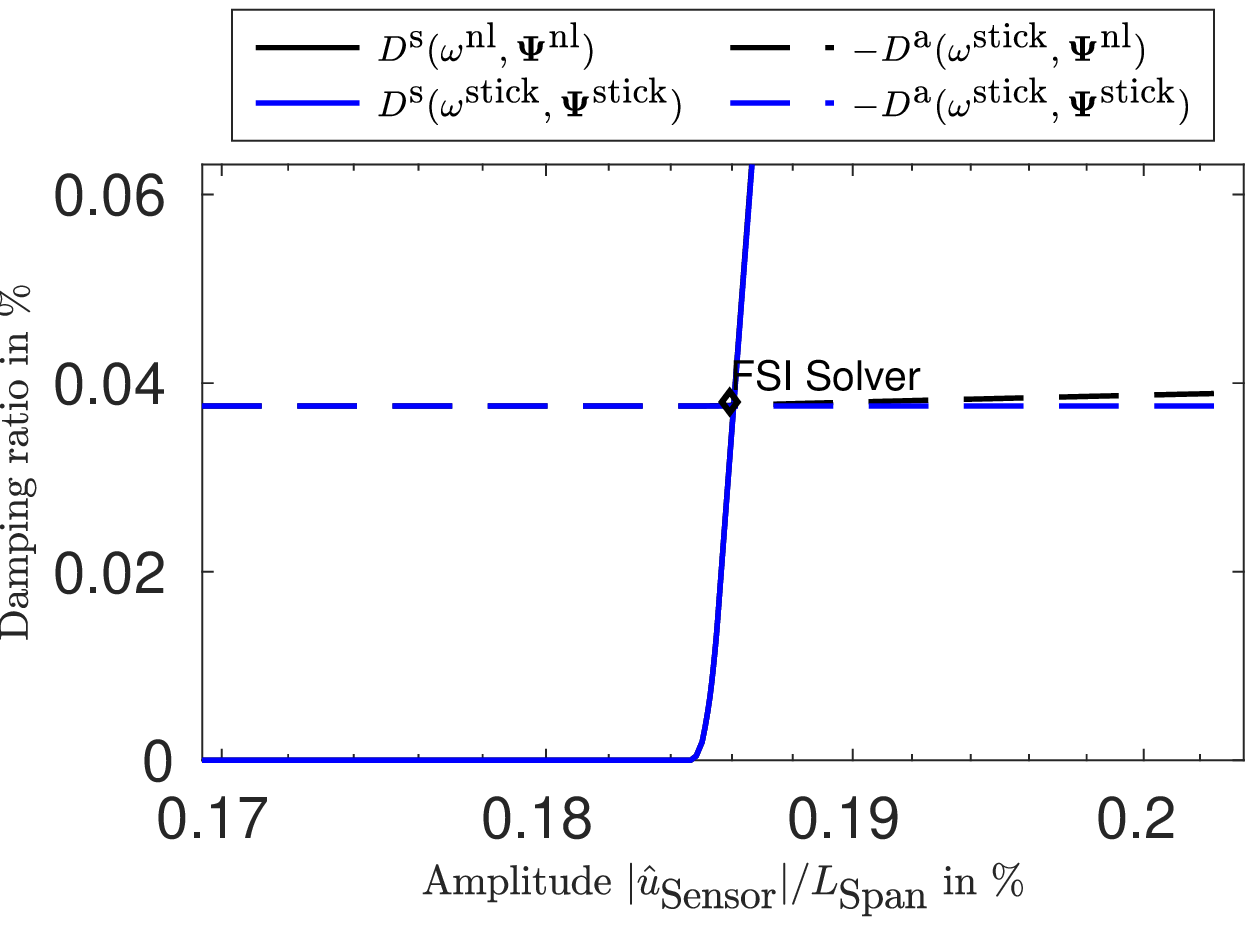}
  \caption{}
  \label{fig:lcoStable/ExcitationDampingVsAmplitudeZoom}
\end{subfigure}
\caption{Results of the conventional and refined energy methods, and the fully-coupled (FSI) solver, configuration 1: \subref{fig:lcoStable/ExcitationDampingVsAmplitude} Global view; \subref{fig:lcoStable/ExcitationDampingVsAmplitudeZoom} Detailed view at LCO solution.}
\label{fig:lcoStableNMA}
\end{figure}

\subsubsection{Numerical performance of the fully-coupled solver}
The decay of the residuals in either domain is shown in \fref{lcoStable/convergenceResiduals} for the first four iterations of the fully-coupled solver.
The residuals decrease with rates typical for the decoupled subproblems.
The convergence of the limit cycle amplitude and frequency is depicted in \fref{lcoStable/convergenceDominantFixedInterfaceModePhase}.
Recall that $\hat{u}^\mathrm{s}_i$ denotes the fundamental Fourier coefficient of the dominant fixed-interface mode.
The results almost do not change after the first iteration and the full-coupled solver reaches convergence in accordance with the specified tolerances already after 5 iterations.
Both the linearization involving the frequency-constant influence coefficient matrix (with the natural frequency of the sticking contact) and the dominant-mode linearization were assessed and found to lead to numerically identical results in the same number of iterations.
The simulation took \num{19.5} hours on 10 cores.
\begin{figure}
\begin{subfigure}{.57\textwidth}
  \centering
  \begin{tikzpicture}
  \def\xstart{13.1}
  \def\yshift{40.5}
  \draw (0, 0) node[inner sep=0] {\includegraphics[width=\linewidth]{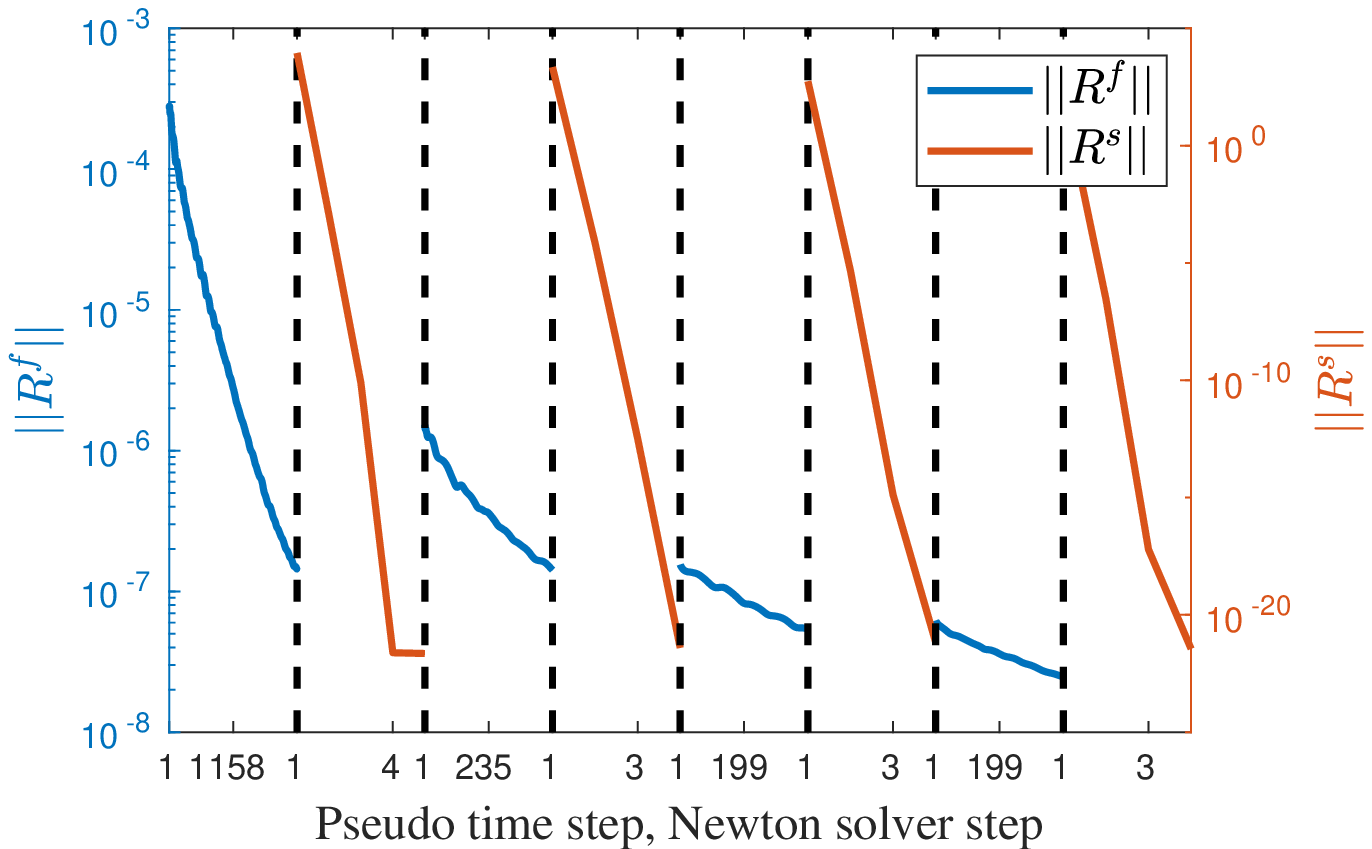}};
  \draw(0,0)node[yshift=\yshift+65]{coupled FSI solver iteration};
  \draw(0,0)node[rotate=90,yscale=5.8,xscale=2,xshift=\yshift,yshift=\xstart-8.67*0](a){\}};
  \draw(0,0)node[rotate=90,yscale=5.8,xscale=2,xshift=\yshift,yshift=\xstart-8.67*1](b){\}};
  \draw(0,0)node[rotate=90,yscale=5.8,xscale=2,xshift=\yshift,yshift=\xstart-8.67*2](c){\}};
  \draw(0,0)node[rotate=90,yscale=5.8,xscale=2,xshift=\yshift,yshift=\xstart-8.67*3](d){\}};
  \draw node[above of=a,yshift=-19]{$m=1$};
  \draw node[above of=b,yshift=-19]{$m=2$};
  \draw node[above of=c,yshift=-19]{$m=3$};
  \draw node[above of=d,yshift=-19]{$m=4$};
  \end{tikzpicture}
  \caption{}
  \label{fig:lcoStable/convergenceResiduals}
\end{subfigure}%
\hspace{0.4cm}
\begin{subfigure}{.37\textwidth}
  \centering
  \includegraphics[width=.99\linewidth]{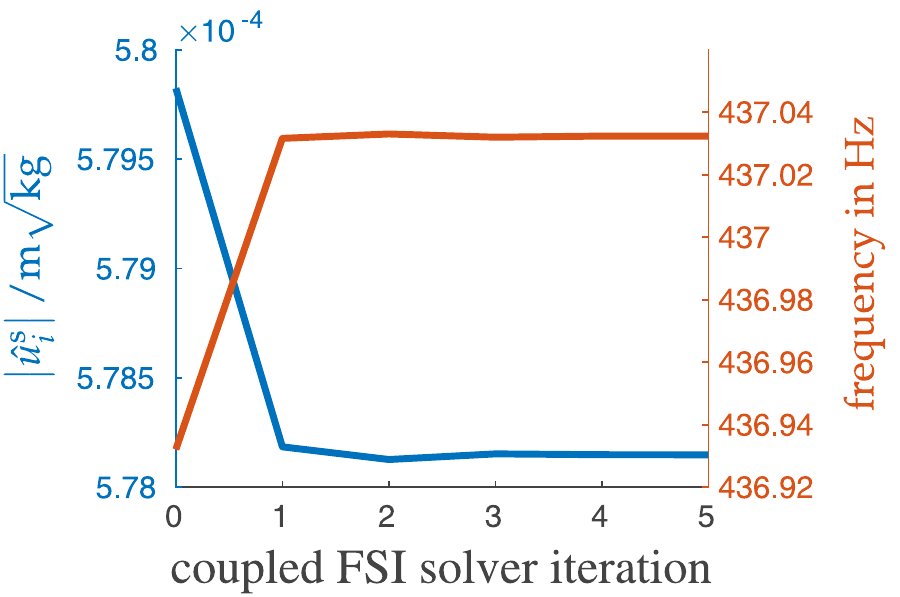}
  \caption{}
  \label{fig:lcoStable/convergenceDominantFixedInterfaceModePhase}
\end{subfigure}
\caption{Behavior of the fully-coupled solver, configuration 1: \subref{fig:lcoUnstable/FlutterCurve} Successive decay of the residuals in fluid and structural domain; \subref{fig:lcoUnstable/FlutterCurveZoom} convergence of the dominant amplitude and frequency of the limit cycle.}
\label{fig:convergenceFSIstableLCO}
\end{figure}
\begin{table}[ht]
\centering
\begin{tabular}[t]{rccc}
\toprule
 & Conventional Energy Method & Refined Energy Method & fully-coupled solver \\
\midrule
Frequency & \SI{436.952}{\hertz} & \SI{436.941}{\hertz} & \SI{437.032}{\hertz}\\
$\usensor / L_\mathrm{Span}$ & \num{1.861e-03} & \num{1.861e-03} & \num{1.859e-03}\\
$MAC(\mm \psi,\mm \psi^\mathrm{stick})$ & \num{1} & \num{0.99999995} & \num{0.9999966} \\
$MAC(\mm \psi,\mm \psi^\mathrm{nl})$ & - & \num{1} & \num{0.9999974}\\
\bottomrule
\end{tabular}
\caption{Summary of stable limit cycle data.}
\label{tab:datastableLCO}
\end{table}

\subsection{Stability Limit (configuration 2)\label{sec:unstableLC}}

\subsubsection{Comparison of the results obtained by energy methods and fully-coupled solver}
The structural and aerodynamic damping ratios are depicted as function of the vibration amplitude in \fref{lcoUnstableNMA}.
In contrast to the case of the stable limit cycle, the unstable limit cycle does not occur close to the linear case with sticking contact conditions.
Consequently, the oscillation frequency and the deflection shape differ considerably from the modal properties of the sticking limit case.
The conventional energy method ignores the amplitude-dependence of the oscillation frequency and deflection shape (assuming $\mm\psi^{\mathrm{stick}}$, $\omega^{\mathrm{stick}}$ for all $\hat u$), and thus the aerodynamic damping ratio is invariant according to this method.
As the damping ratio is positive in the sticking limit case, the conventional energy method does not predict a stability limit, but predicts stable behavior for all amplitudes.
\begin{figure}
\begin{subfigure}{.5\textwidth}
  \centering
  \includegraphics[width=.85\linewidth]{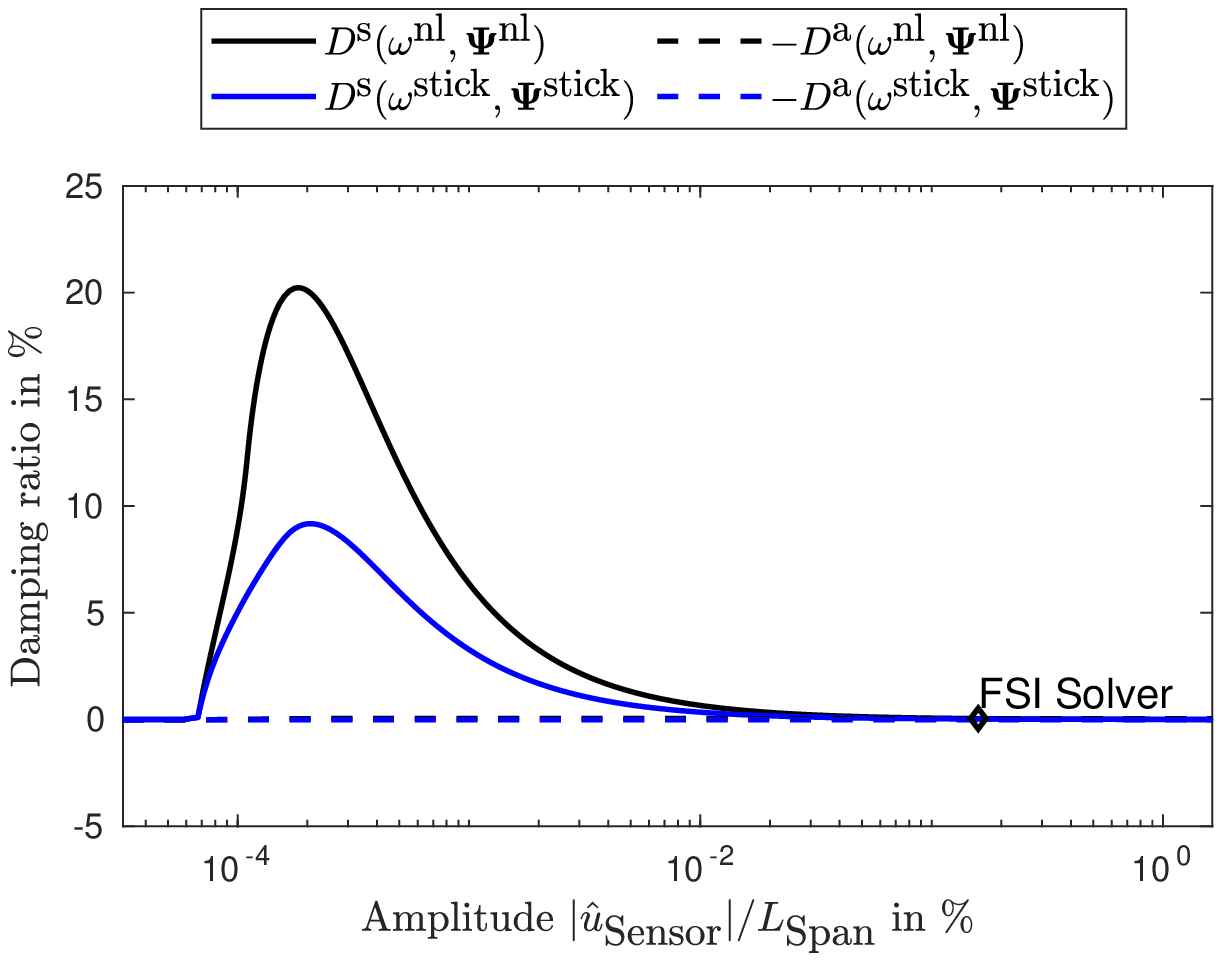}
  \caption{}
  \label{fig:lcoUnstable/ExcitationDampingVsAmplitude}
\end{subfigure}%
\begin{subfigure}{.5\textwidth}
  \centering
  \includegraphics[width=.88\linewidth]{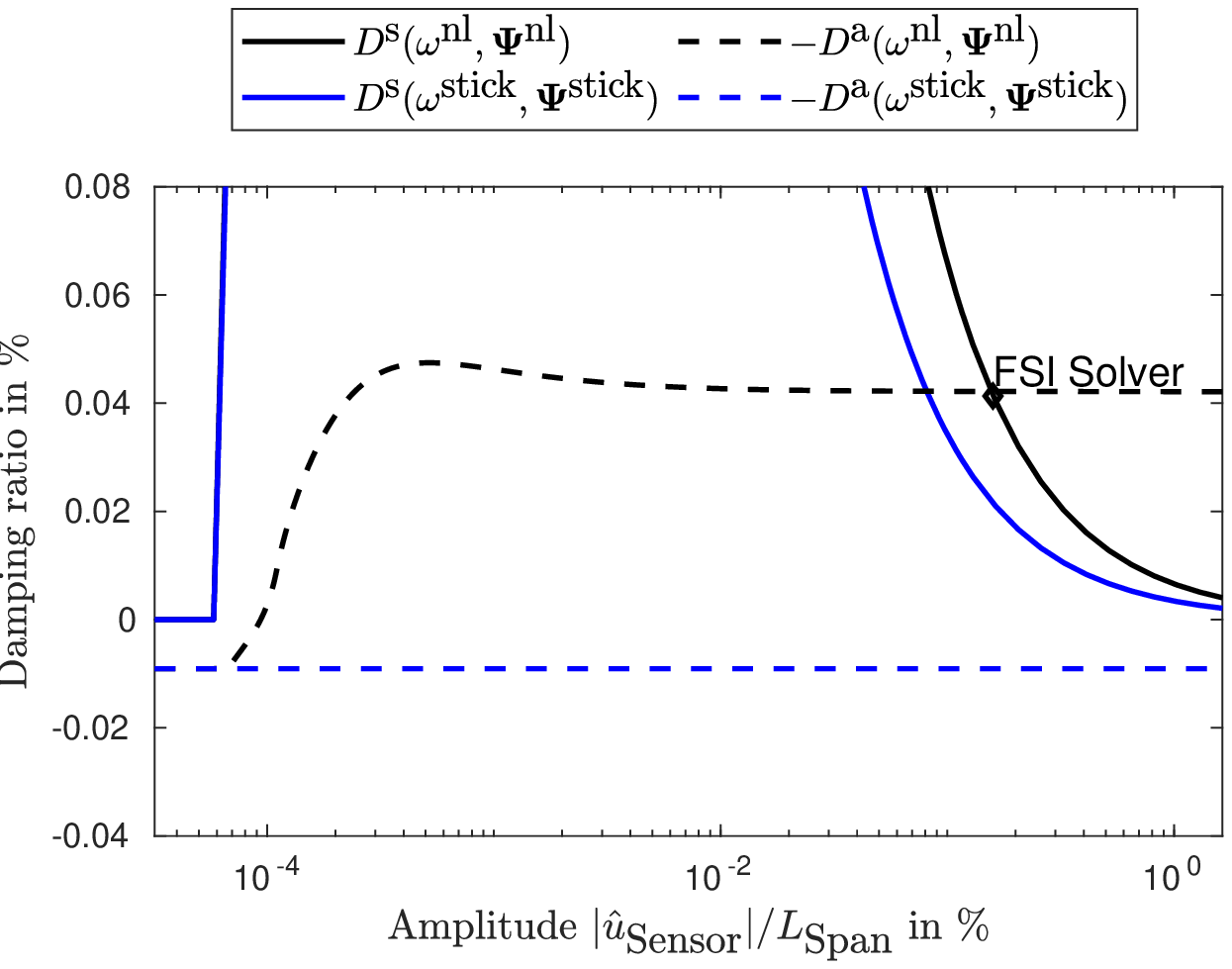}
  \caption{}
  \label{fig:lcoUnstable/ExcitationDampingVsAmplitudeZoom}
\end{subfigure}
\caption{Results of the conventional and refined energy methods, and the fully-coupled (FSI) solver, configuration 2: \subref{fig:lcoUnstable/FlutterCurve} Global view; \subref{fig:lcoUnstable/FlutterCurveZoom} Detailed view.}
\label{fig:lcoUnstableNMA}
\end{figure}
\\
In \tref{dataUnstableLCO} the results of the energy methods and the fully-coupled solver are summarized in terms of amplitude and frequency of the stability limit.
The results of the refined energy method and the fully-coupled solver agree very well.
The deviations of frequency, $\usensor$ are \SI{0.02}{\percent}, \SI{0.88}{\percent} respectively.
The $MAC$ values of about \num{0.91} with respect to the sticking limit case show that the vibrational deflection shape is significantly affected by the nonlinear contact interactions in the tip shrouds.
This leads to a considerable influence on the aerodynamic response, \cf $-\Danma$ and $-\Danmaconstpsi$ in \fref{lcoUnstable/NMAOmegaPsi}.
\begin{table}[ht]
\centering
\begin{tabular}[t]{rccc}
\toprule
 & Conventional Energy Method & Refined Energy Method & full-coupled solver \\
\midrule
Frequency & \SI{519.661}{\hertz} & \SI{418.049}{\hertz} & \SI{418.671}{\hertz}\\
$\usensor / L_\mathrm{Span}$ & - & \num{1.57e-03} & \num{1.59e-03}\\
$MAC(\mm \psi,\mm \psi^\mathrm{stick})$ & \num{1} & \num{0.91069} & \num{0.91073}\\
$MAC(\mm \psi,\mm \psi^\mathrm{nl})$ & - & \num{1} & \num{0.999996}\\
\bottomrule
\end{tabular}
\caption{Summary of stability limit data.}
\label{tab:dataUnstableLCO}
\end{table}
The vibrational deflection shape predicted by the coupled solver and the refined energy method agree very well, as indicated by a $MAC$ value close to unity (right column, bottom row).
This means that the vibrational deflection shape is mainly driven by structural forces, and can thus be well-represented by a nonlinear mode.
Hence, the fluid does not add substantial stiffness (causing a significant frequency shift) or even change the deflection shape.
The aerodynamic self-excitation merely determines the vibration level.

\subsubsection{Influence of the modal deflection shape and frequency}
To further study the influence of the amplitude-dependent modal deflection shape $\mm \psi^\mathrm{nl}$ and frequency $\omega^\mathrm{nl}$, the refined energy method is applied with either the frequency or modal deflection shape held constant, denoted by $-\Danmaconstom$ and $-\Danmaconstpsi$, respectively.
Both results are plotted in \fref{lcoUnstable/NMAOmegaPsi}.
For larger amplitudes, there is a noticeable difference to $-\Danma$ which shows that both effects are important to determine the accurate flow response for this case.
Here, the frequency dependency is mainly responsible for the aerodynamic behavior changing from damping to self-excitation.
This appears plausible since a decrease in vibrational frequency leads to a decrease in reduced frequency which can trigger aerodynamic instability \cite{Srinivasan1997}.
\fig[tbh]{lcoUnstable/NMAOmegaPsi}{Results of the refined energy method (ND -4) with constant frequency or constant modal deflection shape.}{0.4}

\subsubsection{Numerical performance of the fully-coupled solver}
The convergence behavior of the fully-coupled solver is illustrated in \fref{lcoUnstable/convergenceResiduals} and \fref{lcoUnstable/convergenceDominantFixedInterfaceModePhase}.
Again, the solver converges quickly, but more iterations are needed for the same tolerances compared to the previous benchmark configuration.
In this configuration, the type of linearization used for the representation of the aerodynamic forces within the structural subproblem (\cf \sref{flowlin}) was found to have an important influence.
The dominating mode linearization performed well, whereas the frequency-dependent influence-coefficient-matrix linearization did not lead to convergence in a reasonable number of iterations.
The behavior could be improved by introducing a relaxation factor of $0.3$ for $\usolid_{(m)}$, see the results in \fref{lcoUnstable/fullG/convergenceDominantFixedInterfaceModeAmplitude}.
The coupled solver required \num{83.5} hours and \num{62} hours on 10 cores for the  dominating-mode linearization and the influence-coefficient-matrix linearization (with the described relaxation) respectively.
\begin{figure}
\begin{subfigure}{.57\textwidth}
  \centering
  \begin{tikzpicture}
  \def\xstart{13.1}
  \def\yshift{40.5}
  \draw (0, 0) node[inner sep=0] {\includegraphics[width=\linewidth]{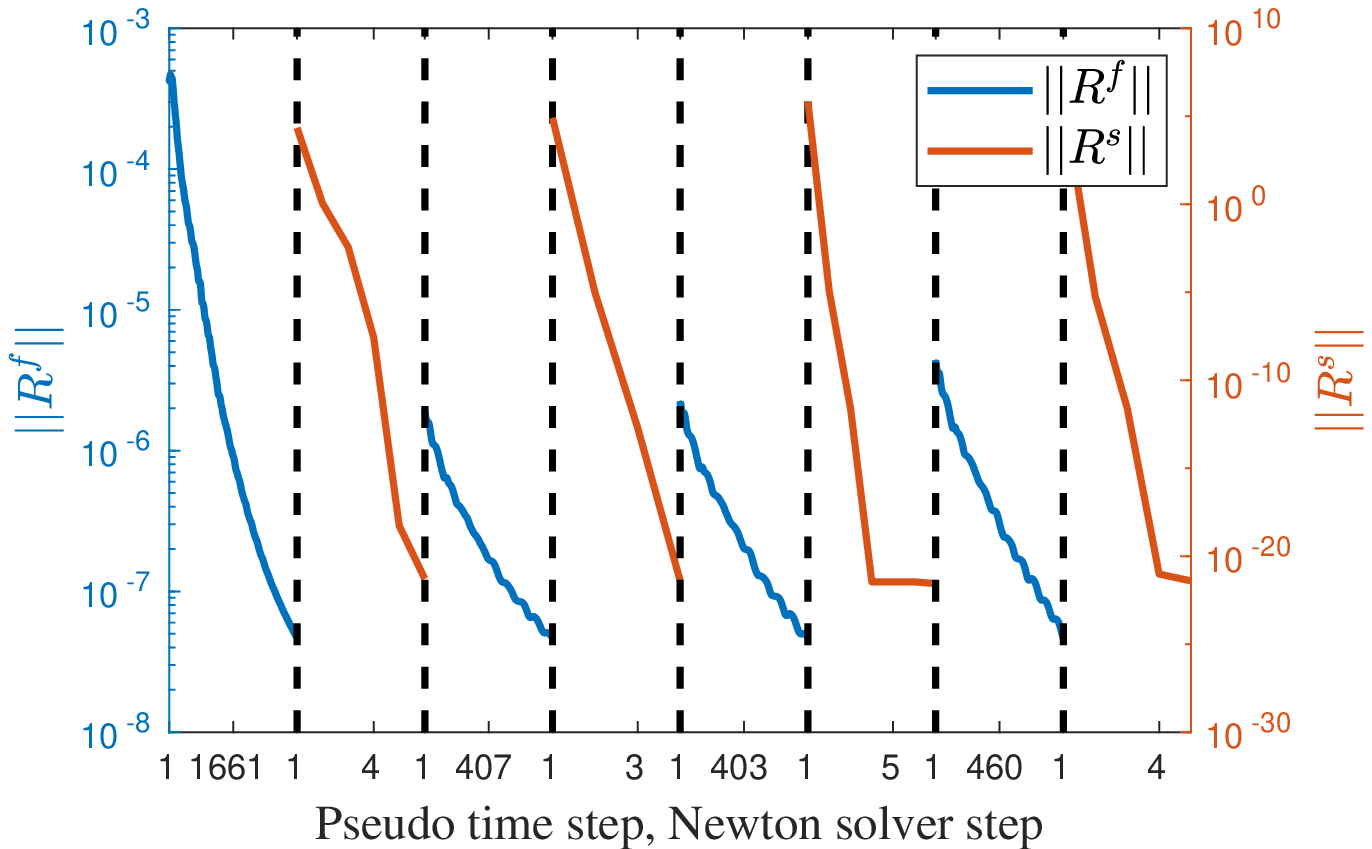}};
  \draw(0,0)node[yshift=\yshift+65]{coupled FSI solver iteration};
  \draw(0,0)node[rotate=90,yscale=5.8,xscale=2,xshift=\yshift,yshift=\xstart-8.67*0](a){\}};
  \draw(0,0)node[rotate=90,yscale=5.8,xscale=2,xshift=\yshift,yshift=\xstart-8.67*1](b){\}};
  \draw(0,0)node[rotate=90,yscale=5.8,xscale=2,xshift=\yshift,yshift=\xstart-8.67*2](c){\}};
  \draw(0,0)node[rotate=90,yscale=5.8,xscale=2,xshift=\yshift,yshift=\xstart-8.67*3](d){\}};
  \draw node[above of=a,yshift=-19]{$m=1$};
  \draw node[above of=b,yshift=-19]{$m=2$};
  \draw node[above of=c,yshift=-19]{$m=3$};
  \draw node[above of=d,yshift=-19]{$m=4$};
  \end{tikzpicture}
  \caption{}
  \label{fig:lcoUnstable/convergenceResiduals}
\end{subfigure}%
\hspace{0.4cm}
\begin{subfigure}{.37\textwidth}
  \centering
  \includegraphics[width=.99\linewidth]{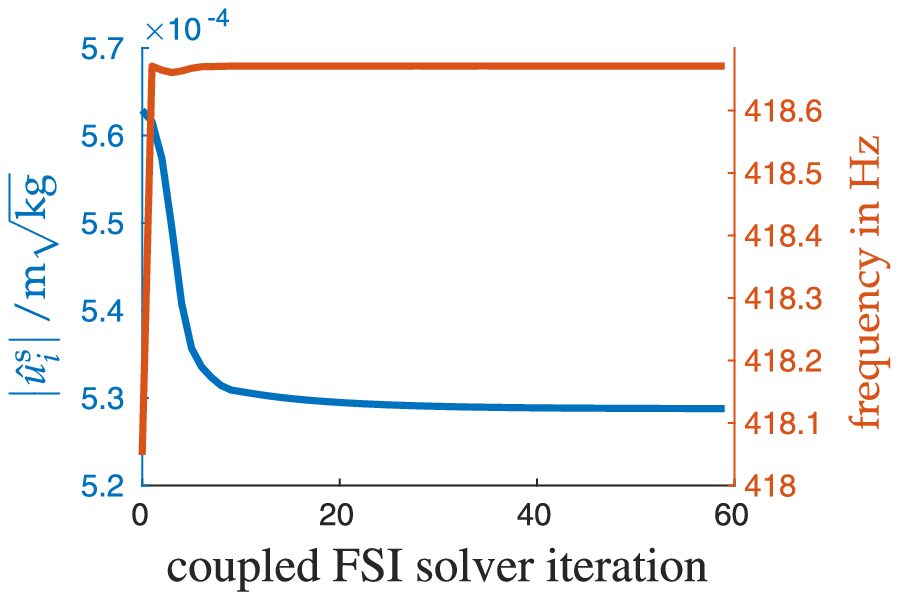}
  \caption{}
  \label{fig:lcoUnstable/convergenceDominantFixedInterfaceModePhase}
\end{subfigure}
\caption{Behavior of the fully-coupled solver, configuration 2: \subref{fig:lcoUnstable/FlutterCurve} Successive decay of the residuals in fluid and structural domain; \subref{fig:lcoUnstable/FlutterCurveZoom} convergence of the dominant amplitude and frequency of the limit cycle.}
\label{fig:convergenceFSIunstableLCO}
\end{figure}

\subsection{Discussion of the computational effort of fully-coupled solver and refined energy method}
The computational costs for solving the structural subproblem was practically negligible in the considered benchmark.
For instance, the nonlinear modal analysis took only about \num{30} seconds.
For the refined energy method, thus the computational effort is driven by the computation of the influence coefficient matrix $\mm G$.
This effort scales with the number of generalized coordinates, $N$, of the reduced structural model (here $N=21$).
For one column of $\mm G$, one flow simulation is required which took approximately \num{5.25} hours on 9 cores.
This sums up to \num{110.25} hours on 9 cores.
For the linear interpolation with respect to frequency, $\mm G$ must be computed for two frequencies, which doubles the computational effort.
The advantage of this approach is that one obtains a closed-form expression for the aerodynamic forces, which can be useful to analyze the aerodynamic damping (and stiffness) for a large number of different vibrational deflection shapes and frequencies (in the considered frequency range).
An alternative to this approach is to directly compute the aerodynamic forces for a given structural vibration, and to determine the resulting aerodynamic damping.
The effort for such a computation is about the same as that for computing a single column of $\mm G$.
Thus, this alternative becomes computationally more attractive than using the influence coefficient matrix once the number of amplitude levels for which $\Delta W^{\mathrm a}$ is of interest falls below the number of generalized coordinates (or twice this number in the case of the two-point interpolation with respect to frequency).

%% file: conclusions.tex
\section{Conclusions\label{sec:conclusions}}
In this work, the conventional energy method has been refined and a fully-coupled frequency-domain method were developed and assessed.
These methods are useful to compute flutter-induced limit cycle oscillations of turbomachinery bladed disks.
In contrast to the conventional energy method, the new methods account for the amplitude-dependence of the oscillation frequency and the vibrational deflection shape.
The refined energy method is particularly suited to separate the different effects of contact and aerodynamic forces, which allows deep understanding of the physics of nonlinear aeroelastic interactions.
As the fully-coupled solver follows a serial coupling strategy, available codes can be used which are optimized for the respective structural and fluid subproblem.
This also permits to take into account various nonlinear effects in either domain in a straight-forward way.
Even if the aerodynamic forces remain linear in the vibration amplitude, the fully-coupled solver might be an attractive alternative.
The reason for this is that the influence coefficients need to be computed for a large number of component modes in order to accurately describe variation of the vibrational deflection shape within the refined energy method.
\\
As benchmark, a state-of-the-art aeroelastic model of a bladed disk subjected to nonlinear contact interactions at the interlocked tip shrouds was considered.
The fully-coupled solver showed excellent convergence behavior for reasonable computational effort.
The good agreement between fully-coupled solver and refined energy method confirms that the aerodynamic forces can, in a wide range, be well approximated as amplitude-linear and using a linear interpolation of the modal aerodynamic influence coefficients with respect to frequency.
In the first configuration, a stable limit cycle under relatively low aerodynamic self-excitation was analyzed.
The oscillation frequency and vibrational deflection shape were thus close to the linear case under sticking contact conditions here.
Hence, the conventional energy method yields a reasonable approximation and is therefore in good agreement with the newly developed methods.
The second configuration demonstrated, for the first time, the phenomenon of nonlinear blade flutter instability, \ie, a situation where the static equilibrium (sticking contact conditions) is stable, but the vibrations start to grow unboundedly once the initial vibration amplitude exceeds a certain threshold.
Neither linear theory nor the conventional energy method are able to predict this behavior, while the newly developed methods are both able to determine the critical amplitude (stability limit).
\\
An open question is under which circumstances a considerable amplitude-nonlinearity of the aerodynamic forces is to be expected and how the fully-coupled solver performs in this case.
It would also be interesting to assess the performance of the methods for a case of more severe aerodynamic influence, where the aerodynamic stiffness is also expected to affect the oscillation frequency and perhaps the vibrational deflection shape.
Finally, the full-coupled solver could be improved with regard to its initialization, as this currently involves the relatively costly computation of the aerodynamic influence coefficient matrix (with respect to the generalized coordinates of the reduced structural model). 

%% file: appendix.tex
\appendix
\setcounter{figure}{0}
\setcounter{table}{0}

\section{Convergence Details} \label{append:convergence}
\fig[ht]{lcoUnstable/fullG/convergenceDominantFixedInterfaceModeAmplitude}{Convergence of the dominant amplitude and frequency of the limit cycle for the sbtability limit with full-influence-matrix linearization.}{0.4} 
%